\documentclass[prd,preprint,superscriptaddress,nofootinbib,showpacs]{revtex4} 

\usepackage{graphicx,amsmath,amsfonts,amssymb,revsymb,dcolumn} 

\newcommand{\be}{\begin{equation}}
\newcommand{\ee}{\end{equation}}
\newcommand{\bea}{\begin{eqnarray}}
\newcommand{\eea}{\end{eqnarray}}

\begin{document}
\title{Electromagnetic vacuum fluctuations and topologically induced  motion
of a charged particle}

\author{C.H.G. \surname{Bessa}} 
\affiliation{Departamento de F\'{\i}sica, Universidade Federal de Campina Grande,
Caixa Postal 5008 \\
58429-900 Campina Grande -- PB, Brazil}
\author{M.J. \surname{Rebou\c{c}as}} 
\affiliation{Centro Brasileiro de Pesquisas F\'{\i}sicas, Rua Dr.\ Xavier Sigaud 150 \\
22290-180 Rio de Janeiro -- RJ, Brazil}


\begin{abstract}
We show that nontrivial topologies of the spatial section  of
Minkowski space-time allow for motion of a charged particle
under quantum vacuum fluctuations of the electromagnetic field. This
is a potentially observable effect of these fluctuations.  
We derive the mean squared velocity dispersion when the charged particle
lies in Minkowski space-time with compact spatial sections in one, two
and/or three directions.
We concretely examine the details of these stochastic motions when
the spatial section is endowed with different globally
homogeneous and inhomogeneous topologies.
We also show that compactification in just one direction of the
spatial section of Minkowski space-time is sufficient to give rise to
velocity dispersion components in the compact and noncompact directions.
The question as to whether these stochastic motions under vacuum
fluctuations can locally be used to unveil global (topological)
homogeneity and inhomogeneity is discussed.
In globally homogeneous space topologically induced velocity dispersion
of a charged particle is the same regardless of the particle's position,
whereas in globally inhomogeneous the time-evolution of the velocity
depends on the particle's position.
Finally, by using the Minkowskian topological limit of globally homogeneous
spaces we show that the greater is the value of the compact topological
length the longer is the time interval within which the velocity dispersion   
of a charged particle is negligible. This means that  no motion of a charged particle
under electromagnetic quantum fluctuations is allowed when Minkowski space-time
is endowed with the simply-connected spatial topology.
The ultimate ground for such stochastic motion of charged particle under
electromagnetic quantum vacuum fluctuations is a nontrivial space topology.
\end{abstract}


\maketitle

\section{Introduction} \label{Intro}

In the framework of general relativity the Universe is modeled as a
four-dimensional differentiable manifold $\mathcal{M}_4$ locally endowed
with a spatially homogeneous and isotropic Friedmann--Lema\^{\i}tre--%
Robertson--Walker (FLRW) metric.
Geometry is a local attribute that gives the intrinsic curvature. Topology is
a global feature related to the compactness and size of a manifold.
In the standard FLRW  model of the Universe the spatial geometry
constrains, but does not dictate the topology of the spatial sections $M_3$
of the space-time manifold $\mathcal{M}_4 = \mathbb{R}\times M_3$.
Thus, two important questions in the FLRW model   
are the geometry and the topology of spatial sections of the   
space-time manifold.

Despite our present-day inability to predict the spatial topology from a fundamental
theory, one should be able to probe it through cosmic microwave background (CMB) or (and)
primordial gravitational waves~\cite{CosmTopReviews,Reboucas2019} observed
recently~\cite{Abbot2016},  which should obey some basic detectability
conditions~\cite{TopDetec}.
For recent topological constraints from CMB data see
Refs.~\cite{Vaudrevange-etal-12,Planck-2015-XVIII}. For some limits of these searches
for topology through a CMB method see Ref.~\cite{Gomero-Mota-Reboucas-2016}.

{}From the field theory viewpoint, in the standard FLRW approach to model the physical world,
besides the points on the space-time manifold, representing the physical events,
there are fields that satisfy appropriate local differential equations (the physical laws).  
Furthermore, it is assumed that the space-time geometries that arise as solutions of
the gravitational field equations constrain the dynamics of these fields. 
However, the topological properties of a manifold precede
its geometrical features and the differential tensor structure with which the gravitation
theories are formulated.
In this way, it is important to determine whether, how and to what extent physical
results on field theory depend upon or are somehow influenced or even driven by
a nontrivial  topology.

Since the net role played by the spatial topology can be better singled out in
the static FLRW space-time, where the dynamical degrees of freedom are frozen, in
this work we shall focus on the static Minkowski space-time, whose spatial
geometry is Euclidean.
The topology of spatial section of the Minkowski space-time
is often taken to be simply-connected infinite Euclidean noncompact manifold $\mathbb{E}^{3}$.
However, the spatial section of Minkowski space-time can also be any one of
the possible $17$ topologically distinct quotient (multiply-connected) manifolds
$\mathbb{E}^3/\Gamma$, where $\Gamma$ is a  discrete group of fixed-point free
isometries or holonomy of $\mathbb{E}^{3}$~\cite{Wolf67,Thurston}. The quotient manifolds
are compact in at least one direction.
The action of $\Gamma$ tiles the covering noncompact manifold $\mathbb{E}^{3}$ into identical
domains or cells which are copies of what is known as fundamental polyhedron (FP) or
fundamental domain or cell (FD or FC).
Hence, the multiply-connectedness (compactness)  gives rise to periodic boundary
conditions (repeated domains or cells in the covering manifold $\mathbb{E}^{3}$) that
are determined by the action of the group $\Gamma$ on $\mathbb{E}^{3}$.

In a  manifold with periodic boundary conditions, only certain modes of quantum
or classical fields are allowed. Hence, the dynamics and the expectation values of
local physical quantities might be affected by the   
global topology. 
Thus, for example, the energy densities for the scalar field in
Minkowski space-time with multiply-connected spatial section
are shifted from  the corresponding values for the Minkowski space-time
with simply-connected spatial topology.
This is a topological Casimir effect and has been investigated
first by DeWitt, Hart and Isham~\cite{dhi79}
(see also Ref.~\cite{Dowker-Critchley-1976})
and more recently by Sutter and Tanaka~\cite{st06}
(Casimir effect of topological origin has also been treated in
Ref.~\cite{MD-2007} and in the related Refs.\cite{DHO-2002,DHO-2001}).

Somehow parallel to this, in recent years there have been fair number of papers
concerning the Brownian motion of test particles subjected to  fluctuations of quantum
fields~\cite{gs99,jr92,wkf02,yf04,yc04,f05,ycw06,hwl08,sw08,sw09,bbf09,bbmm17,pf11,%
whl12,lms14,lrs16}. 
In these articles, the test particles are taken to be classical point particles
coupled with a vacuum fluctuating field.

In the standard Minkowski space-time with simply-connected spatial section,
it is unsettled whether such motions of test particles can occur~\cite{gs99,jr92}.
Thus, these motions under vacuum fluctuations are accomplished through
\textit{changes} in the fluctuations by means of ad hoc insertion of
reflecting plane-boundaries (one or two) into the simply-connected
Minkowski $3-$space. %
In these cases, the mean squared velocity and position of the test particle
are calculated and an effective temperature associated with the transverse
motion is sometimes derived~\cite{yf04,yc04,lms14,lrs16,bbf09,bbmm17}.

The insertion of  plane-boundaries to allow Brownian motions 
from quantum fluctuations of the electromagnetic field in these works, is
ultimately a way of modifying the spatial topology of Minkowski space-time, 
with a resultant $3-$space that is no more a smooth manifold.

A question that naturally arises here is whether conditions that allow vacuum
fluctuations of the electromagnetic field to produce Brownian motion of a
test charged particle can be achieved through nontrivial spatial topologies
of Minkowski space-time. 
In this way, we would have Brownian motions of a test particle in Minkowski
space-time $\mathcal{M}_4$ induced by the nontrivial topology of the spatial
section $M_3$ instead of modifying the spatial topology by the 
introduction of ad hoc plane-boundaries.%
\footnote{The classification of three-dimensional Euclidean spaces was
originally studied in the context of crystallography%
~\cite{Feodoroff-1885,Bieberbach-1911,Bieberbach-1912} 
and was completed in 1934~\cite{Novacki-1934}. For recent
presentation we refer the readers to
Refs.~\cite{Adams-Shapiro01,Cipra02,Riazuelo-et-el03,Fujii-Yoshii-2011}.   }

Our primary aim in this article is to address this question by considering
the motion of test particle with mass $m$ and charge $q$ in Minkowski flat
space-time whose spatial section is endowed with four different nontrivial
flat topologies, namely   
the Slab  ($E_{16}$),  Chimney ($E_{11}$),  Chimney
with half a turn ($E_{12}$), and $3-$Torus ($E_{1})$ topologies.
These topologies turned out to be  
suitable to reveal different topological effects on the motion of test
particles in Minkowski space-time.
\footnote{See next Section for a brief summary on flat three-dimensional topologies,
and Refs.~\cite{Adams-Shapiro01,Cipra02,Riazuelo-et-el03,Fujii-Yoshii-2011}
for more detailed account on these topologies.}

In the next section we set up the notation and present some relevant concepts
and results regarding topologies of three-dimensional manifolds, which
are necessary for the remainder of the paper.
Section~\ref{Stoch-montion} we present our physical systems, the geometric (local)
topological (global) underlying assumptions, derive the eletromagnectic
correlation functions and the velocity dispersion for the motion of a
charged particle under quantum vacuum electromagnetic fluctuations in Minkowski
space-time with spatial sections compact in one, two and three
independent directions.
The question as to whether these stochastic motions under
vacuum fluctuations can be locally used to unveil global (topological)
homogeneity and inhomogeneity is discussed. It emerges
that nontrivial topologies of the spatial section allow for motion of a charged
test particle (potentially observable effect of quantum vacuum fluctuations).
In Section~\ref{Stoch} to further explore the  role played by the spatial
topology we concretely examine the details of the  motion of a charged
particle under electromagnetic fluctuations for the $3-$manifolds endowed
with the globally homogeneous Slab ($E_{16}$),
Chimney ($E_{11}$), $3-$Torus ($E_{1})$, and the globally inhomogeneous
Chimney with half a turn ($E_{12}$) topology.
An important outcome is that compactification in just one direction
is enough to give rise to motion with velocity dispersion components
in the compact and noncompact directions.
In Section~\ref{TopSet} we explain that topological spaces may either be
globally homogeneous or inhomogeneous.    
We show that for the globally homogeneous topologies the topologically
induced effects on the velocity dispersion is the same
regardless of the particle's position. 
Section~\ref{Stoch} we also examine the question as to whether the
motion of a charged particle under vacuum fluctuations can be locally
used to capture or detect the global topological properties of homogeneity
and inhomogeneity of the spatial section of Minkowski space-time.
As it seems unsettled whether electromagnetic quantum vacuum
fluctuations in standard Minkowski space-time with simply-connected
spatial section would allow for motion of a charged particle,  
in Section~\ref{Stoch} we tackle this question
by considering what we call the  \textit{Mikowskian topological limit\/}  of
globally homogeneous manifolds that arises when their compact topological
lengths grow indefinitely. We  found that, regardless the globally
homogeneous topology we start from, the greater is the value of the
topological length, the longer is the time interval in which the  
dispersion is negligible.
The topological Minkowskian limit results make apparent that no motions
of a charged particle arise from electromagnetic quantum fluctuations
in Minkowski space-time with trivial (simply-connected)
topology of the spatial section.

\section{Topological essentials}  \label{TopSet}

The spatial section of Minkowski space-time flat manifold, which is decomposable into
$\mathcal{M}_4 = \mathbb{R}\times M_3$, is often taken to be the simply
connected (noncompact) Euclidean  $3$-manifold $\mathbb{E}^{3}$.
However, they can also be multiply-connected (compact in at least one direction)
quotient manifolds of the form  $M_3=\mathbb{E}^{3}/\Gamma$, where $\mathbb{E}^{3}$
is the covering space, and $\Gamma$ is a discrete and fixed point-free group of discrete
isometries of $\mathbb{E}^{3}$, also referred to as the holonomy
group~\cite{Thurston}.

A simple example of Euclidean (flat) quotient manifold is the so-called Chimney
space, denoted in the literature by $E_{11}$, which is open (noncompact) in one
direction and decomposed into
$E_{11} = \mathbb{R} \times \mathbb{S}^1 \times \mathbb{S}^1  =\mathbb{E}^3/\Gamma$,
where  $\mathbb{R}$ and $\mathbb{S}^1$ stand for the real line and the circle,
respectively.
The  covering space clearly is $\mathbb{E}^3$, and a fundamental domain can
be taken to be an \textit{open parallelepiped} (chimney) with two pairs of opposite
faces identified through translations, together with a noncompact (open) direction
$\mathbb{R}$.
This FD tiles the simply-connected covering space $\mathbb{E}^3$.
The group $\Gamma$ consists of two independent discrete translations associated with
the two face identifications of the FD.
The periodicities in the two independent compact directions are given by the
circles $\mathbb{S}^1$.

The multiply-connectedness of the quotient manifolds $M_3$ leads to periodic
boundary conditions on the covering manifold $\mathbb{E}^3$ (repeated domains
or cells) that are determined by the action of the group $\Gamma$ on the
covering manifold $\mathbb{E}^{3}$. Clearly, different isometry groups
$\Gamma$ define different topologies for $M_3$, which in turn give rise to
different periodicity on the covering manifold.

In addition to the simply connected flat Euclidean space $\mathbb{E}^{3}$,
the spatial section of Minkowski space-time can be any one of the $17$
multiply-connected $3-$dimensional quotient flat manifolds of the form
$\mathbb{E}^{3}/{\Gamma}$.
Assuming orientability of the space section of Minkowski space-time
we have that only nine out of these quotient manifolds are orientable.
They consist of the six compact manifolds, namely $E_1$ ($3-$Torus), $E_2$
(half turn space), $E_3$ (one quarter turn space), $E_4$ (one third turn space),
$E_5$ (one sixth turn space), $E_6$ (Hantzsche-Wendt space), together with
three non-compact, namely the Chimney space $E_{11}$, the Chimney space
with half turn, $E_{12}$, and the Slab space $E_{16}$. For details on the
names and a description of the fundamental domain of these manifolds we
refer the readers to
Refs.~\cite{Adams-Shapiro01,Cipra02,Riazuelo-et-el03,Fujii-Yoshii-2011}.

In Table~\ref{Tb-4-Orient-manifolds} we collect the symbol used to
refer to the manifolds  in which we shall study the motion of
test particles, along with their names and the number of compact
independent directions.

\begin{table}[ht!]
\begin{tabular}{*3{c}}  
 \hline\hline
Symbol   & Name (quotient manifold)      &  Compact dimension \\
\hline
$E_{16}$ &  Slab space                     & $1$  \\
$E_{11}$ &  Chimney space                  & $2$  \\
$E_{12}$ &  Chimney space with half turn   & $2$  \\
$E_1$    &  $\,3\,-$Torus                  & $3$  \\    
\hline\hline
\end{tabular}
\caption{Four out of nine multiply-connected flat orientable quotient
manifolds $M_3 =\mathbb{E}^3/\Gamma$ that are used in the next Sections
in the study of the motion of test particles. The number of compact
dimensions for each manifold is also given.}
\label{Tb-4-Orient-manifolds}
\end{table}

Multiply-connected Euclidean manifolds are not rigid,
in the sense that flat quotient manifolds with the same topology, defined by a
given holonomy group $\Gamma$, can have different sizes. Thus, their topological
compact lengths are not fixed.  In this way, for example, in the three-torus
class of quotient manifolds, the sides $a$, $b$ and $c$ (say) of the faces of the
parallelepiped (fundamental cell) are not fixed (rigid).

An important point regarding the compact flat manifolds $M_3$ is that any
holonomy $\gamma_i \in \Gamma$ of an orientable Euclidean $3-$space can always
be expressed as a screw motion (in the covering space $\mathbb{E}^3$),
which is a combination of a rotation $R(\alpha,\mathbf{\widehat{u}})$
by an angle $\alpha$ around an axis $\mathbf{\widehat{u}}$, say, followed
by a translation along a vector $\mathbf{L} = L \,\mathbf{\widehat{w}}$, say.%
\footnote{The choice of axes to describe a screw motion is not unique.
However, one can always find  a rotation axis parallel to the direction
of the translation vector $\mathbf{L}$.}
The action of a holonomy $\gamma \in \Gamma$ on a generic point $\mathbf{p}$ of
the covering manifold is given by
$\mathbf{p} \rightarrow \gamma \mathbf{p} = R_{}^{}\,\mathbf{p} + \mathbf{L}$.
When there is no rotational part in the screw motion, $\alpha=0$, the
holonomy reduces to a pure translation, and its action
is exactly the same at every point in covering space. In this case, the
distance between $\mathbf{p}$ and its image by the holonomy $\gamma$, namely
$\mid\gamma\,\mathbf{p}-\mathbf{p}\mid = L$, is the   
same for all points $\mathbf{p}$. 
For a proper screw motion ($\alpha \neq 0$), however, the distance
$\mid\gamma\,\mathbf{p}-\mathbf{p}\mid$    
depends on the location of $\mathbf{p}$, and in particular on the distance  
between $\mathbf{p}$ and the axis of rotation.

A way to characterize the shape of compact manifolds is through the size of
their smallest closed geodesics. In more details, for any $\mathbf{x} \in M_3$,
the distance function $\ell_\gamma (\mathbf{x})$ for a given isometry
$\gamma \in \Gamma$ is def\/ined by
\be
\label{dist-function}
\ell_\gamma(\mathbf{x}) = d(\mathbf{x}, \gamma \mathbf{x}) \; ,
\ee
where $d$ is the Euclidean metric defined on $M_3$. The distance function gives
the length of the closed geodesic that passes through $\mathbf{x}$ and is associated
with a holonomy $\gamma$.
In a globally homogeneous manifold the distance
function for any covering holonomy  $\gamma$ is constant. 
However, in globally inhomogeneous manifolds the length of the closed
geodesic associated with at least one  $\gamma$ (non-translational) depends
upon the point $\mathbf{x} \in M_3$, and therefore is not constant.

When the distance between a point $\mathbf{x}$ and its
image $\gamma \mathbf{x}$ (in the covering space) is a constant for all points
$\mathbf{x}$ then the holonomy $\gamma$ is a translation.
Thus, the group elements  $\gamma$'s of the covering
group $\Gamma$  in globally homogeneous spaces are translations.
This means that in these manifolds the faces of the fundamental  
cells are identified through independent translations.
In this way, the above  manifolds $E_{16}$, $E_{11}$ and $E_1$ of
Table~\ref{Tb-4-Orient-manifolds} are globally homogeneous, whereas the
Chimney space with half a turn  $E_{12}$ is globally inhomogeneous since
the covering group $\Gamma$ contains a proper screw motion. 

For  $E_{12}$ there are two generators $(\gamma_t\,, \gamma_s) \in \Gamma$
from which one can derive the expression for a generic $\gamma \in \Gamma$.
Taking the pure translation in the $y-$direction and the rotation axis of the
screw motion parallel to $x$, they are given by
\begin{equation}   \label{gamma_t}
\gamma_t: \;\,
\mathbf{p}  \mapsto \,  \gamma_t \, \mathbf{p} \;\, | \;
\left( \begin {array}{c} x\\ \noalign{\smallskip} y\\ \noalign{\smallskip}   %
z\end {array} \right)
\mapsto \left( \begin {array}{c} x\\ \noalign{\smallskip} y\\ \noalign{\smallskip}
z\end {array} \right)
+  \left( \begin {array}{c} 0\\ \noalign{\smallskip} b\\ \noalign{\smallskip}
0\end {array} \right)\, 
\end{equation}
and the screw motion generator 
\begin{equation}  \label{gamma_s}
\gamma_s: \;\,
\mathbf{p} \mapsto \gamma_s \,\mathbf{p}    \;\, | \;
\left( \begin {array}{c} x\\ \noalign{\smallskip}y\\ \noalign{\smallskip}
z\end {array} \right)
\mapsto  \left( \begin {array}{ccc} 1&0&0\\ \noalign{\smallskip}0&-1&0
\\ \noalign{\smallskip}0&0&-1\end {array} \right)
\left( \begin {array}{c} x\\ \noalign{\smallskip}y\\ \noalign{\smallskip}
z\end {array} \right)
+  \left( \begin {array}{c} a\\ \noalign{\smallskip}0\\ \noalign{\smallskip}
0\end {array} \right)\,,
\end{equation}
where the subindex $t$ and $s$ stand, respectively, for translation
and screw motion, $a$ and $b$ are translational constants, and where
we have used $\alpha = \pi$ (half turn rotation).
Now, clearly the remaining $\gamma_i \in \Gamma$ are given through
successive application of $\gamma_t$ and  $\gamma_s$ to a generic point
$\mathbf{p}$. Thus,  a collective element $\gamma$ can be written
in the form
\begin{equation} \label{gen_gamma}
\gamma : \;  \mathbf{p} \,\, \mapsto \,\, \gamma \mathbf{p}\, \quad | \quad
(x, y, z) \mapsto \Bigl(x + n_x\,a, (-1)^{n_x}\,y + n_y\,b, (-1)^{n_x}\,z\Bigl)\,,
\end{equation}
where $n_x$ and $n_y$ are integers and run from $-\infty$ to $\infty$.
This general expression is displayed later in the Table~\ref{Tb-Spatial-separation}.

Finally, we mention that an interesting outcome of the next section is
that the topological (global) homogeneity and inhomogeneity of the spatial
section of Minkowski space-time are features that can be physically probed 
through the study of the motions of test particles under the influence of
quantum vacuum fluctuations of the electromagnetic field.

\section{Vacuum fluctuations and stochastic motions} \label{Stoch-montion}

Our physical system is a  non-relativistic test particle with
charge $q$ and mass $m$ locally subjected to
vacuum fluctuations of the electric field ${\bf E}({\bf x}, t)$.
It resides in the spatial section $M_3$ of Minkowski space-time
manifold $\mathcal{M}_4$, which is decomposed into
$\mathcal{M}_4 = \mathbb{R}\times M_3$ and is equipped with the
Minkowski metric
\begin{equation} \label{Mink-metric}
ds^2 =\eta_{\mu\nu}\,dx^\mu dx^\nu = dt^2 - dx^2 - dy^2 - dz^2.
\end{equation}
where clearly $\eta_{\mu\nu}=\mbox{diag} (+1, -1, -1, -1)$. In this paper
we use units $\hbar = c =1$.
The topology of the section $M_3$ is often taken to be the simply connected
(noncompact).
Thus, $M_3$ is the Euclidean  $3-$manifold $\mathbb{E}^{3}$. 
Here, however, instead of assuming the simply-connectedness of $M_{3}$, the
spatial section is supposed to be endowed with one of the four orientable
topologies of Table~\ref{Tb-4-Orient-manifolds}.

Regardless of the $3-$space topology,  the interaction between the charged
test particle and the electromagnetic field is locally governed by
Lorentz force law which in the limit of very low velocities reduces to
\begin{equation}\label{eqmotion1}
\frac{d{\bf v}}{dt} = \frac{q}{m} \,{\bf E}({\bf x}, t)\,,
\end{equation}
where  $\mathbf{v} = \mathbf{v}(\mathbf{x}, t)$ is the particle velocity, and
$\mathbf{x}= x\, \mathbf{i} + y \, \mathbf{j} + z \,\mathbf{k}$
is its position at the time $t$. We restrict our study to the case where the 
test particle position $\mathbf{x}$ may be taken constant 
(negligible displacement)~\cite{yf04,lrs16}.

Assuming that the particle is initially  at rest ($t=t_0=0$) the integration of
Eq.~\eqref{eqmotion1} gives
\begin{equation}\label{eqmotion2}
{\bf v}({\bf x}, t) = \frac{q}{m}\int_0^{t}{\bf E}({\bf x}, t)\,dt \,,
\end{equation}
and the mean squared speed in each of the three independent directions
$i = x, y, z$ is given by
%
\footnote{The mean squared velocity fluctuations of the test particle,
hereafter also referred to as
velocity dispersion or simply dispersion is defined by
$\,\langle \,\Delta \mathbf{v}^2(\mathbf{x}, t) \,\rangle \equiv
\langle \, \mathbf{v}(\mathbf{x}, t)\, \mathbf{v}(\mathbf{x}', t) \,\rangle
-\langle \,\mathbf{v}(\mathbf{x}, t) \, \rangle\,\, \langle \,\mathbf{v}
(\mathbf{x}', t)\,\rangle\,$.}
%
\begin{equation}\label{eqdispersion1}
\langle\Delta v^2_i\rangle = \frac{q^2}{m^2} \int_0^t\int_0^t\langle E_i({\bf x}, t'')
E_i({\bf x}', t')\rangle\, dt'' dt'\,,
\end{equation}
where, following Yu and Ford~\cite{yf04}, we have assumed that the electric field can be
expressed as a sum of classical  $\mathbf{E}_c$ and quantum  $\mathbf{E}_q$ parts, such that
only the quantum fluctuations around the mean classical trajectories survive in
the averaging process.
Thus, the correlation function term $\langle E_i({\bf x}, t)E_i({\bf x}', t')\rangle$ in
equation~\eqref{eqdispersion1} corresponds solely to the two-point correlation function of the
quantum part of the electric field $\mathbf{E}= \mathbf{E}_c + \mathbf{E}_q \,$.

Clearly, the  correlation term in the integrand of velocity dispersion  
\eqref{eqdispersion1} can be written in terms of the potential $A_\mu$ of the Maxwell
tensor $F_{\mu\nu} = \partial_\mu A_\nu - \partial_\nu A_\mu\,$. In an appropriate gauge, 
this permits to rewrite the correlation term as a correlation function for a massless scalar field
$D({\bf x}, t;{\bf x}', t')$ that satisfies massless Klein-Gordon equation
$\,\square D({\bf x}, t;{\bf x}', t') = 0\,$. Thus, following Ref.~\cite{bd82}
we have
\begin{equation}\label{eqphi}
\langle A_\mu({\bf x}, t)A_\nu({\bf x}', t')\rangle
= \eta_{\mu\nu}D({\bf x}, t;{\bf x}', t') \,,
\end{equation}
and therefore using the Maxwell tensor $F_{\mu\nu}$ the correlation term of
Eq.~\eqref{eqdispersion1} can be written as
\begin{equation}\label{eqdif-0}
\langle E_i({\bf x}, t)E_i({\bf x}', t')\rangle = \frac{\partial }{\partial x_i}
\frac{\partial}
{\partial {x'}_i}D({\bf x}, t; {\bf x}', t') - \frac{\partial }{\partial t}
\frac{\partial}
{\partial t'}D ({\bf x }, t; {\bf x'}, t') \,.
\end{equation} 

In Minkowski space-time with simply-connected spatial sections ($M_3= \mathbb{E}^3$),
the Hadamard function $D({\bf x}, t;{\bf x}', t')$ is given by~\cite{bd82}
\begin{equation}\label{eqren}
D_0({\bf x}, t; {\bf x}', t') = \frac{1}{4\pi^2(\Delta t^2 - |\Delta \mathbf{x}|^2)} \,,
\end{equation}
where subscript $0$ indicates the simply-connectedness of the spatial section,
$\Delta t = (t - t')$ and  $|\Delta \mathbf{x}| \equiv r $  
is hereafter called  the spatial separation and given by
\begin{equation}\label{eqrE17}
r^2 = (x-x')^2 + (y-y')^2 + (z - z')^2  \,,
\end{equation}
for Minkowski space-time with trivial topology.

However, in Minkowski space-time with nontrivial spatial topology the local Eq.~(\ref{eqren})
holds but instead of the global condition~\eqref{eqrE17} that ensures simply-connectedness
of the spatial section, the \textit{spatial separation} $r$ takes different forms so as to capture
the  periodic conditions, imposed on the covering space $\mathbb{E}^{3}$ by the
covering group  $\Gamma$, which specifies the spatial topology.
Thus, following Ref.~\cite{st06}, in Table~\ref{Tb-Spatial-separation} we collect
the \textit{spatial separations} for the four topologically inequivalent Euclidean spaces we
shall undertake in this paper.%
\footnote{For figures of the fundamental cells along with additional properties of
these three-dimensional Euclidean topologies we refer the readers to
Refs.~\cite{Cipra02,Riazuelo-et-el03,Fujii-Yoshii-2011}.}

It should be noted that for the multiply-connected manifolds the \textit{spatial separation}
of Table~\ref{Tb-Spatial-separation} is in fact the spatial separation for the lift
of two points from the base manifold. The proper spatial separation for a
pair of points in $E_{16}$, for example, is the infimum of the expression given in
Table~\ref{Tb-Spatial-separation}, over all integers $n_x$. However, for the sake of
brevity, throughout the paper we use the simple expression \textit{spatial separation},
for the eparation ($r^2$  given in Table~\ref{Tb-Spatial-separation}) of lifted points
to $\mathbb{E}^{3}$ from the multiply-connected base manifolds $M_3$.

\hspace{-16mm}
\begin{table}[ht!]
\begin{tabular}{*2{l|c}}  
 \hline\hline
Spatial topology       & Spatial separation \\     
\hline
$E_{16}$ - Slab space                    &
$ r^2 = (x - x'- n_x a)^2 + (y - y')^2 + (z - z')^2 $
\\  \hline
$E_{11}$ - Chimney space                  &  \
$r^2= \left( x - x' - n_xa \right)^{2} +\left(  y - y' - n_yb \right)^{2}
+ \left(z - z' \right)^{2}$
\\ \hline
$E_{12}$ - Chimney with half turn   & \
$r^2= \left( x - x' - n_xa \right)^{2} +\left[  y - (-1)^{n_x}y' - n_yb \,\right]^{2}+ \left[
z - (-1)^{n_x}z' \right]^{2} $
\\ \hline
$E_1\,$ -  $\,3\,-$Torus                     &
$r^2 = (x - x' - n_xa)^2 + (y - y' - n_yb)^2 + (z - z' - n_zc)^2 $
\\
\hline \hline
\end{tabular}
\caption{Spatial separation for Hadamard functions of the four multiply-connected flat orientable
quotient manifolds $M_3 =\mathbb{E}^3/\Gamma$. The topological lengths are denoted by $a$, $b$,
$c$, and give the sides of the faces of the fundamental cells. The numbers $n_x$, $n_y$,  $n_z$
are integers and run from $-\infty$ to $\infty$. For each row when all numbers $n_x$, $n_y$,
$n_z$ are simultaneously zero Minkowski spatial separation \eqref{eqrE17} is recovered.  }
\label{Tb-Spatial-separation}
\end{table}

\subsection{CORRELATION FUNCTIONS} \label{subsec-corr}

In this section we give the calculations of the correlation function
of the electric field 
that is required to have the velocity dispersion~\eqref{eqdispersion1} in
Minkowski space-time whose spatial section $M_3$ has a nontrivial topology
with one, two, or three independent compact directions.

We begin by giving the detailed calculations for  the simplest
multiply-connected spatial section that is compact in just one direction
(Table~\ref{Tb-4-Orient-manifolds}), and called Slab space, $E_{16}$.
The fundamental domain of $E_{16}$ is a \textit{double-open parallelepiped}
that is a slab of space with a finite thickness $a$.
{}From Table~\ref{Tb-Spatial-separation} one has that the spatial separation
for  $E_{16}$ is
\begin{equation}\label{eqrE16}   
r^2 = (x - x'- n_x a)^2 + (y - y')^2 + (z - z')^2\,,
\end{equation}
where $a$ is the compact length taken in the direction of the axis $x$ and $n_x$ is
an integer running from $-\infty$ to $+\infty$.
The term $n_x=0$, however, gives rise to an infinity contribution to the
velocity dispersion \eqref{eqdispersion1}.
Indeed, 
for $E_{16}$  topology the function $D({\bf x}, t; {\bf x}', t')$
can be evaluated in terms of $D_0({\bf x}, t; {\bf x}', t')$
through%
\begin{equation}\label{eqD}     %
D({\bf x}, t; {\bf x}', t') = {\sum}_{n_x=-\infty}^\infty D_0({\bf x}, t;
{\bf x}' + n_x\,a\, {\bf i }, t')      
 = {\sum}_{n_x=-\infty}^\infty \, \frac{1}{4\pi^2(\Delta t^2 - r^2)}\,.
\end{equation}

For all  multiply-connected manifolds the calculations are similarly
made in the covering space endowed with periodic conditions dictated by the covering
group $\Gamma$, which specifies each spatial topology. Thus, in general one
has~\cite{dhi79,st06,Dowker-Critchley-1976,BGRT-1998,GRTB-2000,MMS-2015}
\begin{equation} \label{DsD0}
D({\bf x}, t; {\bf x}', t') = \sum_{\gamma_i \in \Gamma}^{}
\; D_0({\bf x}, t; \gamma_i {\bf x}, t') \,,
\end{equation}
where $D_0({\bf x}, t; {\bf x}', t')$ is the Hadamard function for the
simply-connected $3$-space.%
\footnote{The idea behind this procedure is that as we identity
${\bf x} \sim {\bf x}'$ then for a generic field $\varphi_a ({\bf x},t)$
one has $\varphi_a ({\bf x}) = \varphi_a (\gamma {\bf x})$,
where the subindex $a$ is a generic index (spacetime or internal index).
Thus, for periodic conditions on the covering space one has the sums as
in the above equations~\eqref{DsD0} and \eqref{eqD} (for more details see
Refs.~\cite{dhi79,st06,Dowker-Critchley-1976,BGRT-1998,GRTB-2000,MMS-2015}).
This is called the method of images and is often used in harmonic analysis
~\cite{Terras-2013}.}

Returning to $E_{16}$ case inserting Eq.~\eqref{eqD} into  Eq.~\eqref{eqphi}
and then  into (subsequently)
Eq.~\eqref{eqdispersion1} one finds an infinity contribution due to the
term $n_x = 0$, which has to be subtracted from Eq.~\eqref{eqD}.
The renormalized version of the correlation function is then given by
\begin{equation}\label{eqDren}
D_{ren}({\bf x}, t; {\bf x}', t')  = {\sum}_{n_x=-\infty}^{'\infty}
\;\; \frac{1}{4\pi^2(\Delta t^2 - r^2)}\,.
\end{equation} 
where here and in what follows $\sum_{}^{\;'}$ indicates that the Minkowski
contribution term $n_x = 0$ is excluded from the summation.

We note that the general functional form \eqref{eqDren} of
$D_{ren}({\bf x}, t; {\bf x}', t')$ holds for the four topologies of
Table~\ref{Tb-Spatial-separation}.
However, extra summation terms are added to the left hand side of Eq.~\eqref{eqDren}
depending on the number of compact dimensions of the spatial sections $M_3$.
Thus, for the four multiply connected spatial sections $M_3$ one can use
Eqs.~\eqref{eqphi} and~\eqref{eqDren} to have the renormalized version of
\eqref{eqdif-0}, namely
\begin{equation}\label{eqdif}
\langle E_i({\bf x }, t)E_i({\bf x}', t')\rangle = \frac{\partial }{\partial x_i}
\frac{\partial}
{\partial {x'}_i}D_{ren}({\bf x}, t; {\bf x'}, t') - \frac{\partial }{\partial t}
\frac{\partial}
{\partial t'}D_{ren}({\bf x}, t; {\bf x'}, t') \,.
\end{equation} 
Obviously, as $r$ changes with the topology (see Eq.~\eqref{eqDren}) so does the
function $D({\bf x}, t; {\bf x'}, t')$.

For the Slab space, $E_{16}$, the spatial separation $r$ is given by Eq.~\eqref{eqrE16}
and the electric field correlation function \eqref{eqdif} reduces
to
\begin{equation}\label{electric}
\langle E_i({\bf x}, t)E_{i}({\bf {x'}}, t')\rangle = {\sum}_{n_x=-\infty}^{'\infty} \,\,\,
\frac{\Delta t^2 + \widetilde{r_i}^2}{\pi^2\left[\Delta t^2 - r^2\right]^3}\,,
\end{equation}
where $\widetilde{r_i}^2 = r^2 - 2(\Delta x_i - n_{x_i}a)^2$, for $i= x, y, z $, with
$\Delta x_i = x_i - x_i'$.  
In other words, $\widetilde{r_x}^2 =  r ^2 - 2(\Delta x - n_{x}a)^2$,      
$\,\widetilde{r_y}^2 = r^2 - 2(\Delta y)^2$ and $\,\widetilde{r_z}^2 = r^2 - 2(\Delta z)^2$
for $E_{16}$, in which clearly the compact lengths $b=c=0$.

An analogous procedure can be employed to have the electric field correlation function
for the multiply-connected spatial topologies with two independent compact directions,
namely $E_{11}$ (Chimney space) and $E_{12}$ (Chimney with half turn space).
Indeed, using the spatial separation $r$ for each of these two topologies
(Table~\ref{Tb-Spatial-separation}) along with Eq.~\eqref{eqdif}  
one obtains the general form
\begin{equation}\label{electric2}
\langle E_i({\bf x}, t)E_i({\bf x'}, t')\rangle =
\frac{1}{\pi^2B_i}{\sum}_{n_y=-\infty}^{'\infty}
\; \left(  \, \, {\sum}_{n_x=-\infty}^{'\infty}
\left[ \frac{A_i\Delta t^2 + \widetilde{r_i}^2}{(\Delta t^2 - r^2)^3}\right]
\,\,\right) \,,
\end{equation}
which holds for either spaces $E_{11}$ or $E_{12}$. Clearly, both $\widetilde{r_i}$
as well as the dimensionless constants $A_i$ and $B_i$ depend on each topological
space and on the direction $i = x, y, z$. Thus, for example, for chimney space with
half turn $(E_{12})$, in the coincidence limit ($\mathbf{x} \rightarrow  \mathbf{x}'$)
one has
$A_x=B_x=1$, $A_y=A_z = 3-(-1)^{n_x}$ and $ B_y= B_z = 2$.

Finally, for multiply-connected topologies with three independent compact spatial
directions a similar procedure furnishes the general form
\begin{equation}\label{electric3}
\langle E_i({\bf x}, t)E_i({\bf x'}, t')\rangle =\frac{1}{\pi^2B_i}{\sum}_{n_z=-\infty}^{'\infty}
\; \left[  \, \, {\sum}_{n_y=-\infty}^{'\infty}
\;  \left(  \, \,  {\sum}_{n_x=-\infty}^{'\infty}\left[ \frac{A_i\Delta t^2 +\widetilde{r_i}^2}
{(\Delta t^2 - r^2)^3}
\right]   \,\,\right) \,\,\right] \,,
\end{equation}
where again $\widetilde{r_i}$ and the dimensionless constants $A_i$ and $B_i$ depend on
the underlying topological space and on the direction $i = x, y, z$.

\subsection{VELOCITY DISPERSION} \label{subsec-dispers}

In this section we shall derive general expressions for the velocity dispersion
[ Eq.~\eqref{eqdispersion1} ] for the four multiply-connected spatial spaces.
To this end, we consider separately the manifolds of Table~\ref{Tb-4-Orient-manifolds}
with one, two and three compact directions and the associated correlation function
calculated in the previous section. 
For the case with just one compact dimension ($E_{16}$), from Eqs.~\eqref{eqdispersion1}
and~\eqref{electric} one has that the velocity dispersion 
can be written as
\begin{equation}
\langle \Delta v_i^2\rangle = \frac{q^2}{\pi^2m^2}\int_0^t\int_0^t dtdt' \,
{\sum}_{n_x=-\infty}^{'\infty} \left[ \frac{\Delta t^2
+ \widetilde{r_i}^2}{(\Delta t^2 - r^2)^3}\right].
\end{equation}
Integrating the right hand side of this equation we obtain the general form
\begin{equation} \label{eqI1I2}
\langle \Delta v_i^2\rangle = \frac{q^2}{\pi^2m^2}\,{\sum}_{n_x=-\infty}^{'\infty}
                         \left(\, I + J \, \right) \,,    
\end{equation}
where
\begin{eqnarray} \label{Int1}
\!\!\!\, I = \widetilde{r_i}^2 \int_0^t\int_0^t dtdt' \left[ \frac{1}{(\Delta t^2
- r^2)^3}\right] = \!\!
\frac{-\widetilde{r_i}^2t}{16r^5(t^2-r^2)}\left\{-4rt + 3(r^2-t^2)\ln\left[\frac{(r-t)^2}{(r+t)^2}
\right]\right\}\!,
\end{eqnarray}
and
\begin{eqnarray} \label{Int2}
\!\!\!\,\,J = \int_0^t\int_0^t dtdt' \left[ \frac{\Delta t^2}{(\Delta t^2 - r^2)^3}\right] =
\frac{-t}{16r^3(t^2-r^2)}\left\{-4rt - (r^2-t^2)\ln\left[\frac{(r-t)^2}{(r+t)^2}\right]\right\}\,.
\end{eqnarray}
For the compactification in the direction $\mathbf{i}$, inserting, respectively,
Eqs.~\eqref{Int1} and~\eqref{Int2} into  Eq.~\eqref{eqI1I2} after some simplifying
calculations one obtains
\begin{eqnarray}\label{eqdispersion2}
\!\!\!\langle \Delta v_i^2\rangle =\!\! \sum_{n_x=
-\infty}^{\infty\;\,'}\frac{t\,q^2}{16\pi^2m^2r^5(t^2-r^2)}
\left\{\!4rt(\widetilde{r_i}^2+r^2)   + (t^2 - r^2)(3\widetilde{r_i}^2 - r^2)
\ln\left[\frac{(r-t)^2}{(r+t)^2}
\right]       \right\}\!.
\end{eqnarray}

Following an analogous procedure, i.e. using Eqs.~\eqref{eqdispersion1} and
\eqref{electric2}
[or  \eqref{electric3}] for the spatial section with two 
and three  compact dimensions, after some simplifying calculations  
one obtains
\begin{align}\label{eqdispersion3}
\langle \Delta v_i^2\rangle
&= \sum\limits_{{\!n_y=-\infty}}^{{\infty\;\;'}}    
\biggl[ \,\,  \sum\limits_{{n_x=-\infty}}^{{\infty\;\;'}}         
\frac{t\,q^2}{16\pi^2m^2B_ir^5(t^2-r^2)}  \biggr. \nonumber \\ 
& \biggl. \biggl. \qquad \times \; \left\{4\,rt(\widetilde{r_i}^2+A_ir^2)
+(t^2 - r^2)(3\widetilde{r_i}^2 - A_ir^2) \ln\left[\frac{(r-t)^2}{(r+t)^2}\right]
\right\} \,\,\,\,   \biggr] \,,
\end{align}
and
\begin{align}\label{eqdispersion4}
\langle \Delta v_i^2\rangle
&= \sum\limits_{{\!n_z=-\infty}}^{{\infty\;\;'}}    
\biggl(\,\,\,\sum\limits_{{n_y=-\infty}}^{{\infty\;\;'}}         
\biggl[\,\sum\limits_{{n_x=-\infty}}^{{\infty\;\;'}}         
\frac{t\,q^2}{16\pi^2m^2B_ir^5(t^2-r^2)}  \biggr. \biggr. \nonumber \\ 
&\qquad \times \; \left\{4\,rt(\widetilde{r_i}^2+A_ir^2) +(t^2 - r^2)(3\widetilde{r_i}^2
- A_ir^2) \ln\left[\frac{(r-t)^2}{(r+t)^2}\right]  \right\} \,\,\biggr]   \,\,\,\,  \biggr) \,,
\end{align}
respectively.

In brief, equations~\eqref{eqdispersion2}~--~\eqref{eqdispersion4} give
the velocity dispersion for the four spatial sections $M_3$ of Minkowski space-time
$\mathcal{M}_4  = \mathbb{R}\times M_3$ with nontrivial orientable topologies
with one, two and three compact independent directions, respectively.
The spatial separations $r$ vary with the specific spatial topology.
The term $\widetilde{r_i}$ as well as the constants $A_i$ and $B_i$
change with the spatial topology and with the direction ($i = x, y, z$).

\section{Topologically induced stochastic motions}  \label{Stoch}

Equations~\eqref{eqdispersion2}, \eqref{eqdispersion3} and \eqref{eqdispersion4} of
previous section give the general expressions for the  velocity dispersion of
massive charged test particles in Minkowski space-time with nontrivial spatial
topologies with one, two and three compact dimensions.  Here we concretely study
the details of the  topologically induced stochastic motions for the spatial
orientable topologies of Table~\ref{Tb-4-Orient-manifolds}.

\subsection{\bf Slab space -- $\mathbf{E_{16}}$  } \label{secSlab}

We begin by considering the simplest nontrivial Euclidean spatial topology, namely
Slab space ($E_{16}$). For the compact direction in the direction of the axis $x$,
from Eq.~\eqref{electric} with $r$ given by Eq.~\eqref{eqrE16}, in the coincidence limit
$\mathbf{x} \rightarrow \mathbf{x}'$ 
one has   $r^2 = \widetilde{r_y}^2 = \widetilde{r_z}^2 = -\widetilde{r_x}^2 = n_x^2 a^2$.
This leads to the following components of the correlation function:
\begin{equation}\label{correlationE16x}
\langle E_x({\bf x}, t)E_x({\bf x}, t')\rangle =
{\sum}_{n_x=-\infty}^{'\infty} \;\; \frac{1}{\pi^2[\Delta t^2 - n_x^2a^2]^2}
\qquad \mbox{and} \qquad 
\end{equation}
\begin{equation}\label{correlationE16y}
\langle E_y({\bf x}, t)E_y({\bf x}, t')\rangle
= \langle E_z({\bf x}, t)E_z({\bf x}, t'\rangle =
{\sum}_{n_x=-\infty}^{'\infty} \;\;
\frac{\Delta t^2 + n_x^2a^2}{\pi^2[\Delta t^2 - n_x^2a^2]^3}\,.
\end{equation}

The corresponding  components of the velocity 
dispersion~\eqref{eqdispersion2}
reduce then to
\begin{equation} \label{dispersionxs1}
\langle\Delta v^2_x\rangle = \frac{-q^2t}{4\pi^2m^2}
{\sum}_{n_x=-\infty}^{'\infty} \;\;    
\frac{1}{n_x^3a^3}\ln\left[\frac{(n_xa - t)^2}{(n_xa+t)^2}\right]
\qquad \mbox{and} \qquad 
\end{equation}
\begin{equation} \label{dispersionys1}
\langle\Delta v^2_y\rangle = \langle\Delta v^2_z\rangle = \frac{q^2t}{8\pi^2m^2}
{\sum}_{n_x=-\infty}^{'\infty} \;\;    
\left\{\frac{4t}{n_x^2a^2(t^2-n_x^2a^2)}+\frac{1}{n_x^3a^3}
\ln\left[\frac{(n_xa - t)^2}{(n_xa+t)^2}\right]\right\}\,.
\end{equation}

\begin{figure*}[tb]
\begin{center} 
\includegraphics[width=7.1cm,height=5.9cm]{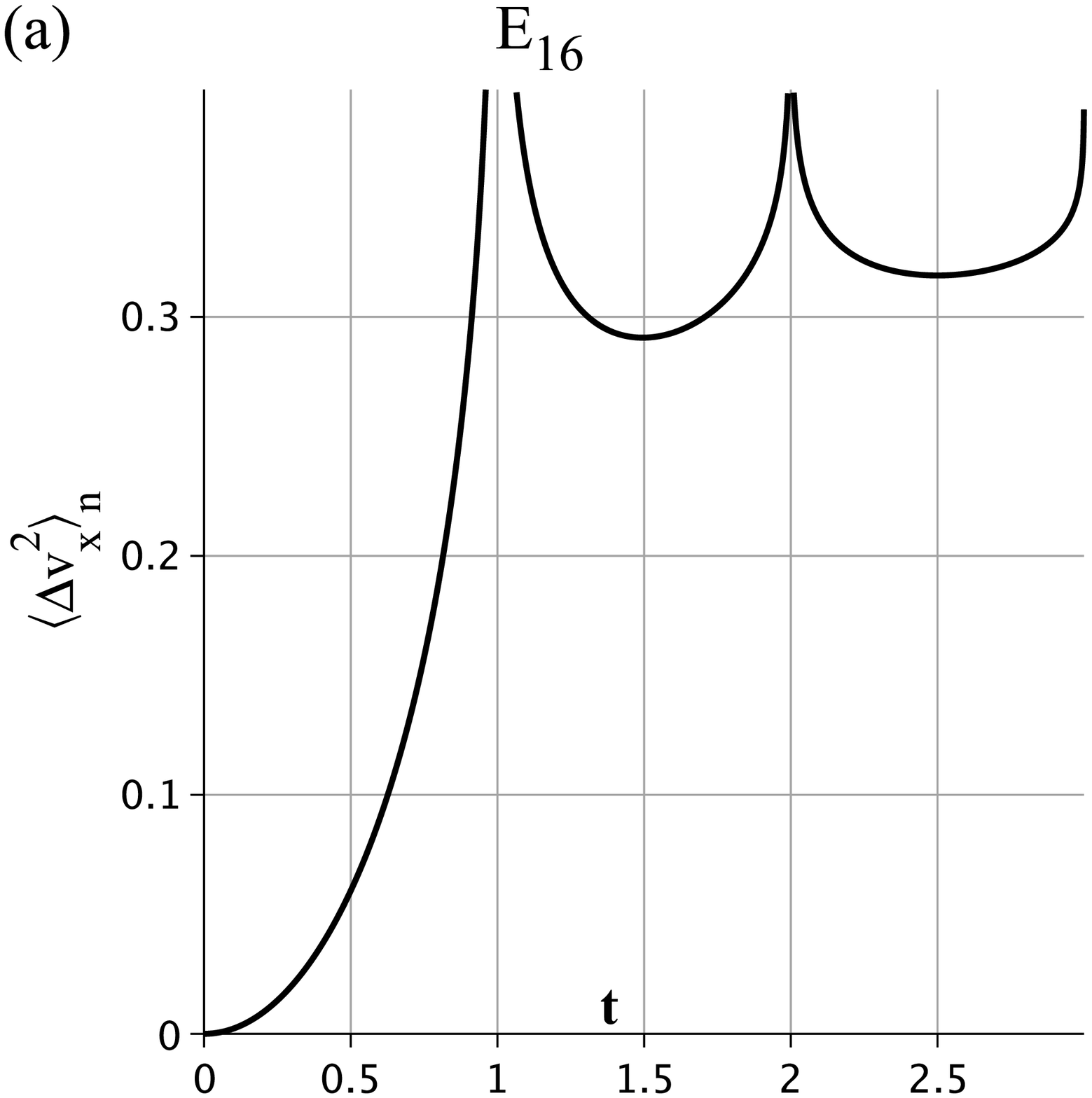}    
 \hspace{8mm}  %
\includegraphics[width=7.1cm,height=5.9cm]{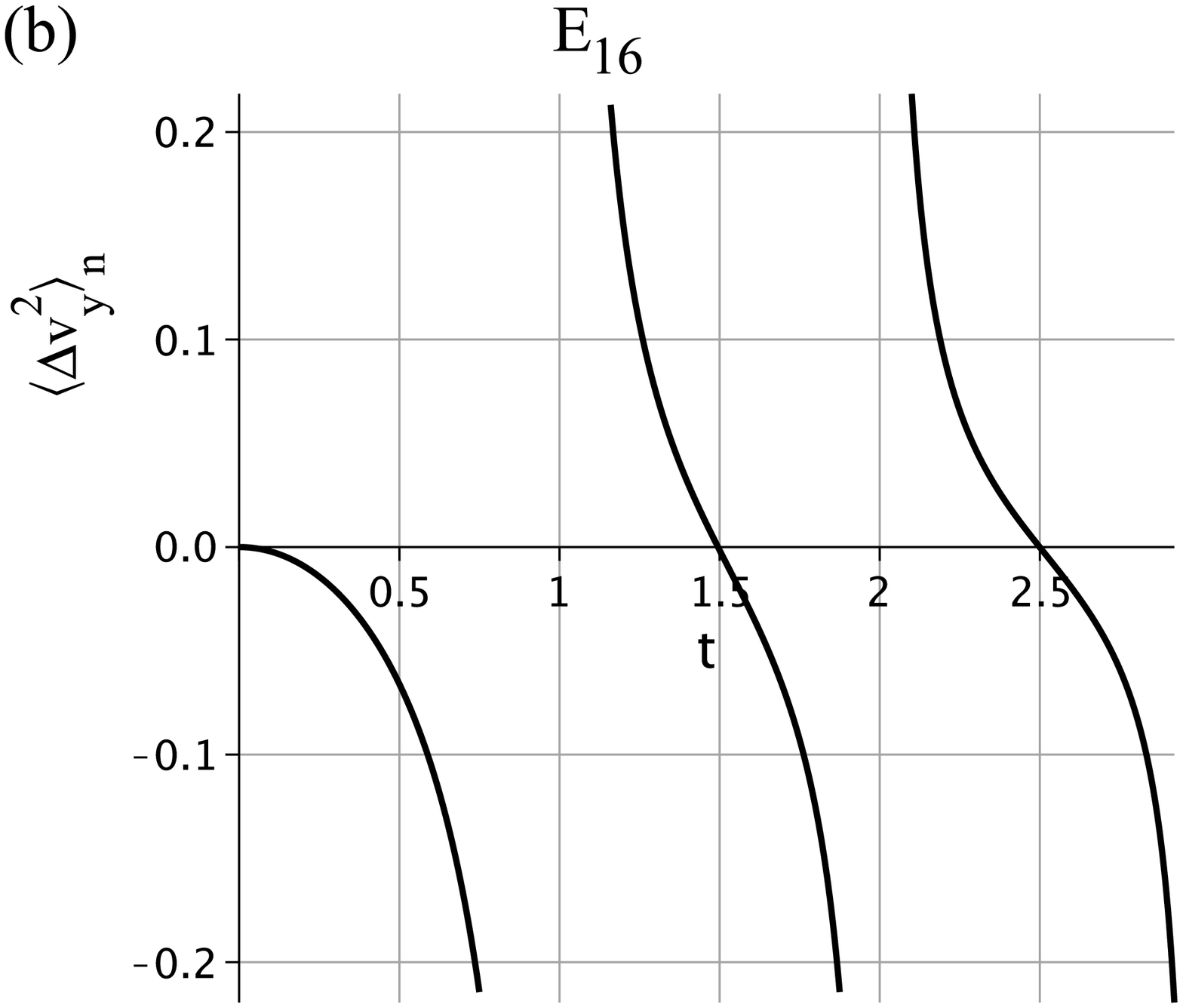}  
\begin{minipage}[t]{\textwidth} \vspace{-5mm} \renewcommand{\baselinestretch}{0.96}
\caption{The $x$  and  $y$  components [panels ({\bf a}), left and ({\bf b}), right)]
of the normalized velocity dispersion $\langle{\Delta} \mathbf{v}^2
(\mathbf{x}, t) \rangle_{\,\mbox{n}}$ [Eq.~\eqref{normalized-disp}]
of a test particle with mass $m$ and charge $q$ in Minkowski space-time
with spatial section endowed with the nontrivial $E_{16}$ topology
with compact length $a=1$.
This figure make apparent that the nontrivial $E_{16}$ topology
allows for motion of charged particle under electromagnetic vacuum 
fluctuations. Clearly, from the topological symmetry the curve for
$z$ component is similar to the curves for $y$ component of the
velocity  dispersion.
The divergent behavior of the normalized velocity
dispersion components for some values of the time $t$ arises from periodic
conditions  imposed by $E_{16}$ topology on  
the covering space $\mathbb{E}^3$.
\label{E_16xy-fig1-ab}  }  
\end{minipage}
\end{center}
\end{figure*}

A first important outcome from equations~\eqref{dispersionxs1}
and~\eqref{dispersionys1}, which is also made apparent through the
panels ({\bf a}) and ({\bf b}) of  Fig.~\ref{E_16xy-fig1-ab},
is that compactification in one direction gives rise to stochastic
motion of charged particles. Moreover, the resulting dispersion
$\langle \Delta \mathbf{v}^2 \left(\mathbf{x}, t\right) \rangle$ has
components not only in the compact direction but also
in the two other independent noncompact directions $y$ and $z$
with the same functional form~\eqref{dispersionys1}.

Panels ({\bf a}) and ({\bf b}) of Fig.~\ref{E_16xy-fig1-ab} show,      
respectively, the time behavior of the  $x$ and $y$ components of the
normalized mean squared velocity fluctuations 
\begin{equation}  \label{normalized-disp}
\langle{\Delta} \mathbf{v}^2 (\mathbf{x}, t) \rangle_{\,\mbox{n}} \equiv
(m^2/q^2)\,\langle\Delta \mathbf{v}^2\,(\mathbf{x}, t )\rangle\,.
\end{equation}
In the numerical calculations to plot the curves in both panels of
Fig.~\ref{E_16xy-fig1-ab} we have set the compact length $a=1$,
and taken $100$ terms of topological contribution to the dispersion,
namely $n_x \neq 0$ in the interval $[-50,50]$.

From equations~\eqref{dispersionxs1} and \eqref{dispersionys1} %
one clearly has divergent behaviors of the normalized velocity dispersion
[Eq.~\eqref{normalized-disp}], which are shown in both panels of
Fig.~\ref{E_16xy-fig1-ab}. These divergent behaviors arise from the
periodic conditions ($ x \rightarrow  x' = x + n_x a$) imposed by $E_{16}$
topology on the covering space.
The singular points of the dispersion  correspond to  $t = n_x a\,$
for $\,n_x = 1, 2$ and $3$. The periodicities are given by circles
$\mathbb{S}^1$, and the singular points correspond to the periods of
time required for light to  travel, respectively, $\,n_x = 1, 2$ and $3$
times the compact length $a$.
For all nontrivial topologies we are concerned with in this paper
the covering manifold is infinite and the covering group  $\Gamma$ tiles
$\mathbb{E}^{3}$ into infinite cells. Therefore, there are infinite
singularities of topological origin that arise at certain discrete
times $t$ and are associated to periodic conditions imposed by the
underlying spatial topology. We briefly indicate below a possible
approach to regularize these singularities.
In what follows in the figures of the dispersion we exhibit only
two (sometimes three or just one) of such singularities, though.

Similar divergent behavior has been reported in the study of the velocity
dispersion for the motion of charged particles under electromagnetic
fluctuations in Minkowski empty space, whose motions are obtained via what
can be looked upon as a \textit{change in the spatial topology} through the
insertion of a plane into  the simply-connected $3-$space section of 
Minkowski $4-$D  space-time~\cite{yf04,yc04}.  
The insertion of a  plane along with the rigid perfectly reflecting boundary
conditions (method of images~\cite{BrowMaclay-1969}) can then be seen
as a topological intervention that tiles the covering noncompact $3-$space
into two identical domains, and the resulting spatial section is not a
manifold, but  the nontrivial inhomogeneous quotient orbifold that 
arises from the action of $\mathbb{Z}_2$ on $\mathbb{E}^3$ by reflection
(see example 13.1.1 in Chapter~13 of Thurston book~\cite{Thurston}). Thus, the
half-space is a quotient space. 
The motion of the charged particle depends on the particle's position
relative to the plane.
The important point is that one can look upon the particles motion 
as induced by the nontrivial spatial topology, which allows for the motion
of test particle under electromagnetic fluctuations in the Minkowski space-time.
An important circle $\mathbb{S}^1$ in this inhomogeneous $3-$space is defined
through the double of this distance from a point to the plane boundary (round-trip).
This double-distance plays similar role to that of the compact length $a$
in the Slab topology.  But, unlike the case with the Slab topology in which
we have infinite singular points, with the insertion of perfectly reflecting
plane boundary we have two identical domains with contribution for the
dispersion, and with just  one singular point for the whole time evolution
of the dispersion.%
\footnote{Perfectly reflecting (absorbs no energy) plane is a physical way to mimic in
a local lab experiment the $3$--space topology of the reflection orbifold.
Its contribution to the motion    
of the particle is to set a nontrivial  topology for spatial sections
of Minkowski space-time, and thus to allow for the particle's motion
under electromagnetic fluctuations.}

Assuming that the divergence of the velocity dispersion is result of
simplified hypothesis about the physical system, \textit{local} modification
through the introduction of a switching function connecting different
stages of the system has been suggested to
regularize the divergence~\cite{sw08,sw09,lrs16,lr19}.
The  ad hoc parameters of the switching functions are thought of as 
local characteristic of the original system such as width and duration
of the switching process, for example.
Nevertheless, for the simple topology as $E_{16}$ the compact length $a$ and
the time $t$ are equal up to a constant, and thus the proposed switching
function can be recast in terms of the topological parameters $a$ (and $n_x$)
to similarly regularize the divergences taking into account its topological
origin.
It should be noted that for topologies with higher degree of
multiply-connectedness, it is necessary a function of topological
parameters, or perhaps several functions, to locally regularize all the
divergences of topological origin. This is an issue beyond the scope
of the present paper, though.

Returning to Fig.~\ref{E_16xy-fig1-ab} from  panel {\bf a} of one has
that the component of dispersion along the compact direction is positive
during the whole time evolution.
This feature permits the definition of an effective temperature associated
to the component of the stochastic motions as in previous works,  where
plane boundaries have been used~\cite{yf04,yc04}.
Panel {\bf b} on the other hand,  exhibits negative dispersions in some
interval of time. Negative dispersion has also been reported in previous works
with both original~\cite{yf04} and modified~\cite{yc04,lms14,lrs16}  system
and has been often intrepreted as a consequence of the renormalization with
respect to the Minkowski vacuum fluctuations.
In Sec.~\ref{subsec-corr} we have also subtracted the Minkowski vacuum fluctuations
term by excluding $n_x = 0 $ from the summation in equation~\eqref{eqDren},
with similar effect in the eletromagnetic correlation function for the Slab and other
topologies considered in this paper.

Differently from the motion obtained via the insertion of ad hoc planes into
the trivial Minkowski $3-$space~\cite{yf04,yc04}, which breaks its
smooth manifold attributes as well as it global spatial homogeneity, here the
motion of test particle under vacuum fluctuations takes place in Minkowski
space-time whose spatial section is a globally homogeneous
smooth manifold $M_3$ with $E_{16}$ topology.  
Furthermore, with insertion of planes the dispersion evolution depends on
the position of the test particle relative to the plate.
Here, however,  since
$E_{16}$ is globally homogenous the dynamical behavior of the particles
is independent of its position $\mathbf{x}$ on the spatial section of
Minkowski space-time.

Two important limiting cases of the above velocity dispersion components \eqref{dispersionxs1}
and \eqref{dispersionys1} are for $a \gg c\,t$ and for $a \ll c\,t$, where $a$ is the compact
length and here $c$ is the speed of light, which has been taken equal to $1$ throughout this
paper.
The former limit can be obtained either topologically, by increasing the compact length $a$
for a fixed finite period of time, or by fixing a given topological length $a$ (fix a
manifold) and restricting locally the time evolution of the particle to a fixed finite
small time $t$. 

For a sufficiently large compact fixed length $a$, which defines a spatial manifold, and for
small time $t$ ($\,t \ll a\,$), from equations~\eqref{dispersionxs1}~--~\eqref{dispersionys1}
after some simplifying calculations one locally has
\begin{equation}\label{resultvx1}
\langle\Delta v_x^2\rangle \approx \frac{\pi^2q^2}{45m^2}\frac{t^2}{a^4}\,,
\end{equation}
\begin{equation}\label{resultvz1}
\langle\Delta v_y^2\rangle = \langle \Delta v_z^2 \rangle
                        \approx -7\times 10^{-2} \frac{q^2t^4}{m^2a^6}\,.
\end{equation}

Panels ({\bf a}) and ({\bf b}) of Fig.~\ref{E_16xy-fig1-ab} also illustrate the different
behavior in this limit for small time and signs of the dispersions in $x$ and $y$ (or $z$)
directions in this limit. Equations \eqref{resultvx1} and \eqref{resultvz1} along with these
panels show the absolute values of mean squared velocity fluctuations tends distinctively
to zero. The nontrivial topology plays different role even for small time in compact and
noncompact directions.

Throughout this paper we shall refer to the limits obtained by increasing the
topological length, such as $a \gg t $,  as Minkowskian topological limit
(or simply Minkowskian limit), in the sense that the role played
by the topology when the compact length, $a$, grows 
tends to that of the simply-connected spatial topology.%
\footnote{Clearly the limiting process of increasing the compact length $a$
is permitted because the topological length in Euclidean quotient manifolds are
not fixed (constant) and therefore it can take different values without changing
the topology. Different compact lengths $a$ correspond to  different $3$-manifolds
with the same topology, though.}
As the topological compact length tends to infinity the mean squared velocity
fluctuations in the noncompact directions $y$ and $z$ decreases to zero
faster than the component in the compact direction $x$.

\begin{figure*}[tb]
\begin{center} 
\includegraphics[width=7.1cm,height=5.9cm]{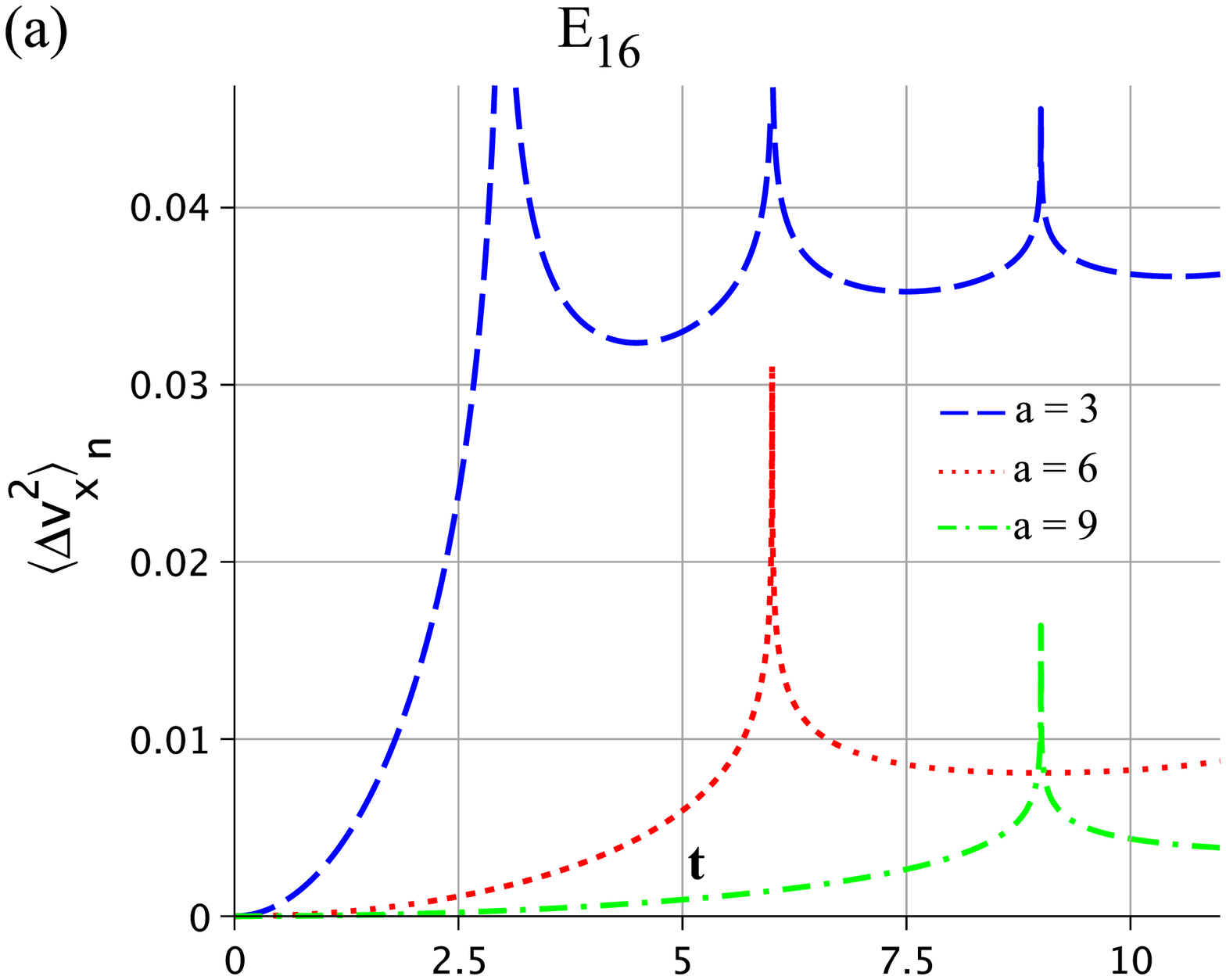}  
 \hspace{8mm}
\includegraphics[width=7.1cm,height=5.9cm]{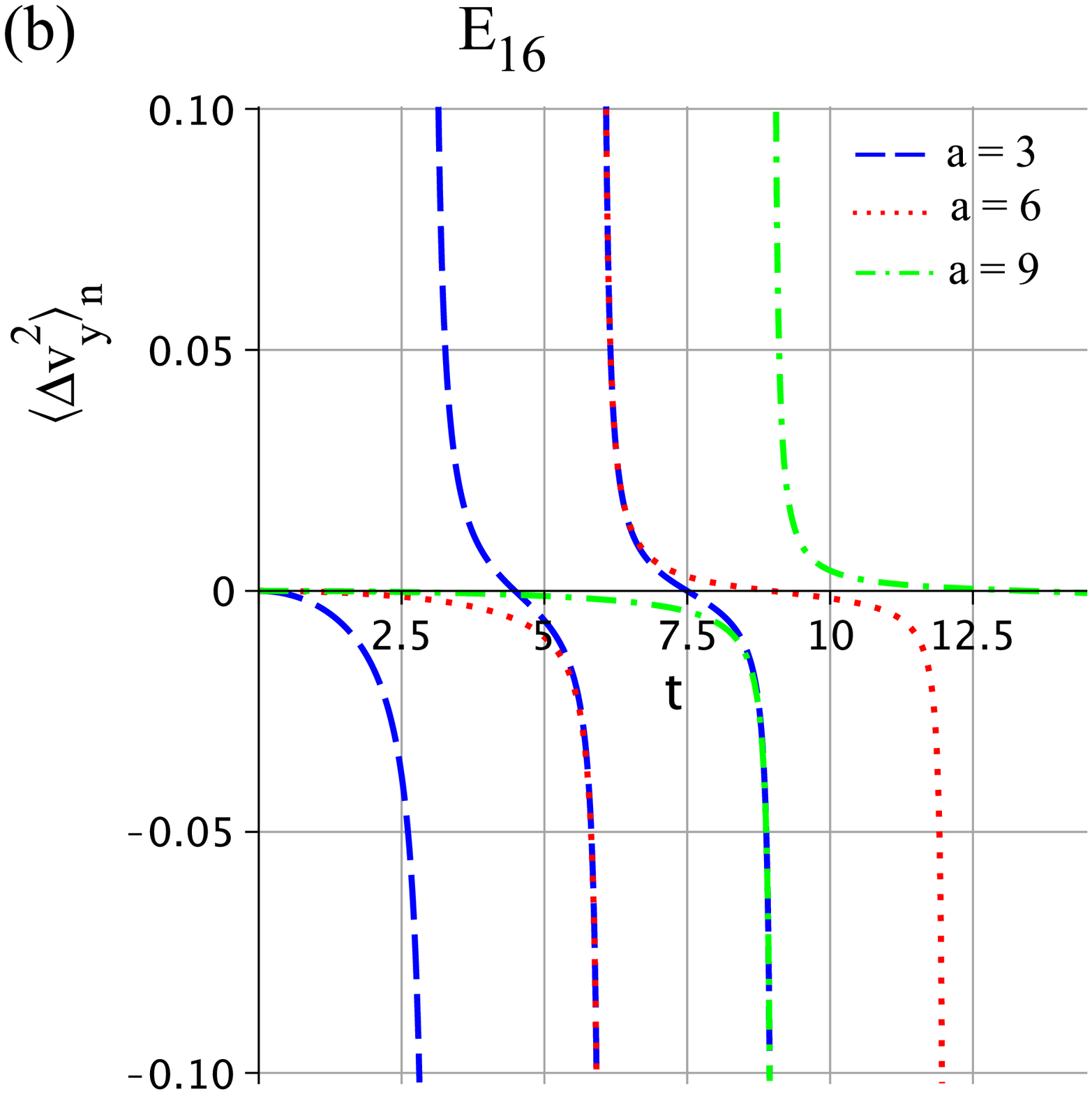}  
\begin{minipage}[t]{\textwidth} \vspace{-5mm}  \renewcommand{\baselinestretch}{0.96}
\caption{[Panel ({\bf a}), left]: $x$  and [panel ({\bf b}), right]: $y$  components
of the normalized velocity dispersion
$\langle{\Delta} \mathbf{v}^2
(\mathbf{x}, t) \rangle_{\,\mbox{n}}$ \  [Eq.~\eqref{normalized-disp}]
of a $q$-charged  test particle with mass $m$ in Minkowski space-time
with nontrivial $E_{16}$ spatial topology.
The curves for manifolds with increasing  $a = 3, 6, 9 $  are shown
to illustrate the topological Minkowskian limit when the compact length
$a$ increases. For these curves the greater is the value of the compact
topological lengths the longer is the time interval $\Delta t_{\mbox{ng}}$
for which the velocity dispersion of is negligible.
In the topological Minkowskian limit $a \rightarrow \infty $ one
has $\Delta t_{\mbox{ng}} \rightarrow \infty $, and thus during the
whole time there is no motion of a charged particle in
Minkowski space-time with simply-connected spatial topology.
The topological symmetries of $E_{16}$ topology imply that
curves analogous to those for $y$ component hold for the $z$ component
of the normalized velocity dispersion.
\label{E16xy_Minkows}  } 
\end{minipage}
\end{center}
\end{figure*}

Panels ({\bf a}) and ({\bf b}) of Fig.~\ref{E16xy_Minkows} illustrate the topological
Minkowskian limit as the compact length  takes different values $a = 3, 6, 9 $.
For these curves the greater is the value of the compact topological length $a$
the longer is the time interval $\Delta t_{\mbox{ng}}$ for which the velocity
dispersion is negligible.
Thus, in the  Minkowskian topological limit $a \rightarrow \infty$ the whole
space is simply-connected and  $\Delta t_{\mbox{ng}} \rightarrow \infty $,
i.e. during the whole time there is no velocity dispersion 
of the charged particle in Minkowski space-time with simply-connected spatial
topology.
This outcome  agrees with earlier result which claims that in the
standard simply-connected Minkowski space-time  vacuum quantum fluctuations
would not produce stochastic motions of charged particles~\cite{gs99,jr92}.
We shall return to the topological Minkowskian limit in the next section where we
investigate the motions of test particle in Minkowski space-time with spatial section
endowed with topologies with higher degree of connectedness such as Chimney and
$3-$Torus spaces, for example.

\subsection{\bf  Chimney space -- $\mathbf{E_{11}}$ } \label{secChimney}

\begin{figure*}[tb]
\begin{center} 
\includegraphics[width=7.1cm,height=5.9cm]{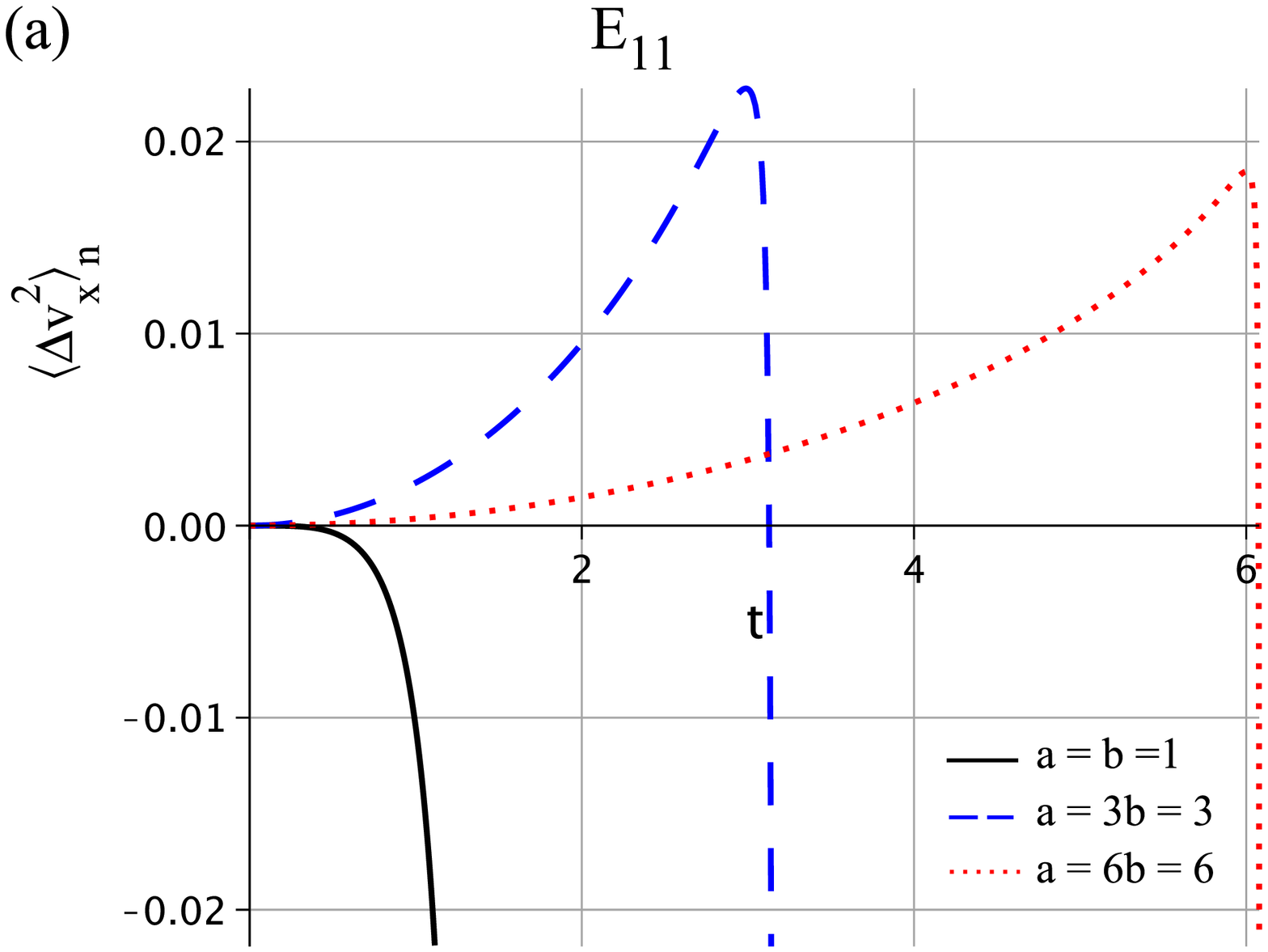}   
 \hspace{8mm}  %
\includegraphics[width=7.1cm,height=5.9cm]{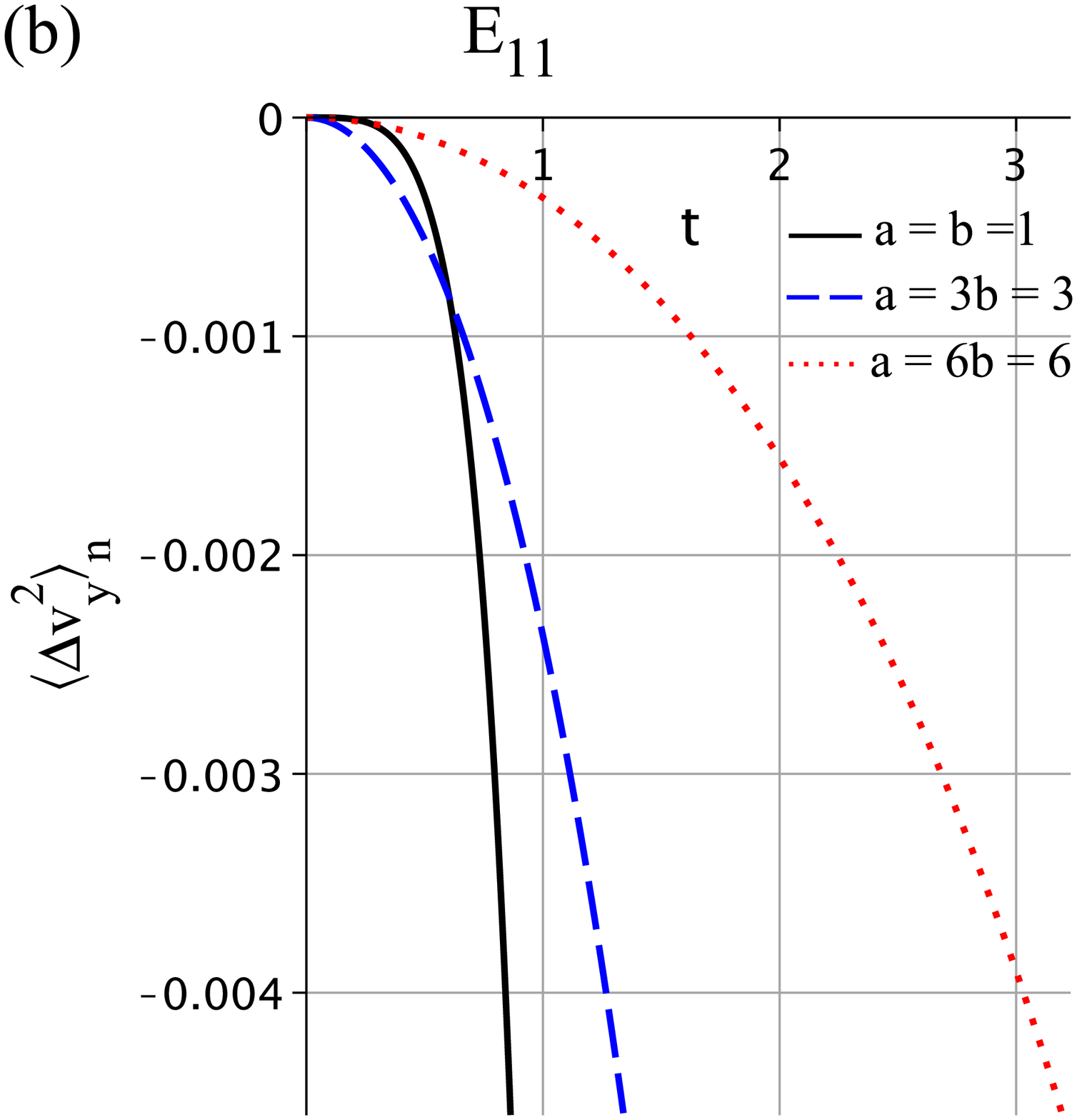}   
\begin{minipage}[t]{\textwidth}  \renewcommand{\baselinestretch}{0.96}  
\caption{The $x$  and  $y$  components [ panels ({\bf a}) and ({\bf b}) ]
of the normalized velocity dispersion
$\langle{\Delta} \mathbf{v}^2
(\mathbf{x}, t) \rangle_{\,\mbox{n}}$ \  [see Eq.~\eqref{normalized-disp}]
particle with mass $m$ and charge $q$ in Minkowski space-time whose spatial
has the nontrivial Chimney topology $E_{11}$. Curves for manifolds  with
fixed $b=1$ and three increasing compact lengths $a=1, 3, 6$ are displayed
to illustrate the dispersion in topological Minkowskian limit
for $E_{11}$ topology.
For these curves the greater is the value of the compact
topological lengths the longer is the time interval $\Delta t_{\mbox{ng}}$
for which the velocity dispersion of is negligible.
In the topological Minkowskian limit $a \rightarrow \infty $ the whole space
is simply-connected, and one has $\Delta t_{\mbox{ng}} \rightarrow \infty $,
so during the whole time there is no velocity dispersion 
of the charged particle in Minkowski space-time with simply-connected spatial
topology.
From the topological symmetries, $z$ is a  noncompact direction with distinct
time evolution and separately shown  in Fig.~\ref{E_11z-a124}. 
\label{E_11xy-a124}  }.
\end{minipage}
\end{center}
\end{figure*}

\begin{figure}[bt]
\begin{center} 
\includegraphics[width=7.1cm,height=5.9cm]{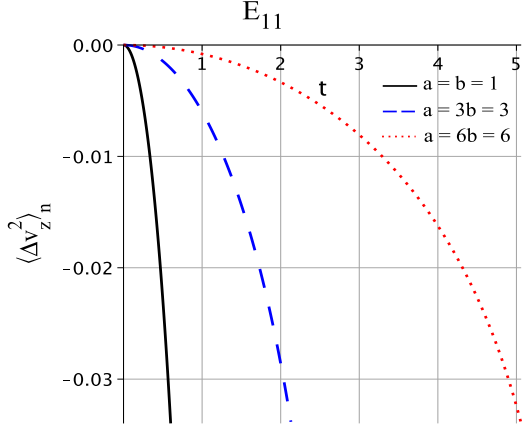}  
\begin{minipage}[t]{\textwidth}  \renewcommand{\baselinestretch}{0.96}
\caption{
The $z$  component of the normalized velocity dispersion
$\langle{\Delta} \mathbf{v}^2 (\mathbf{x}, t) \rangle_{\,\mbox{n}}$ \
[\ Eq.~\eqref{normalized-disp}]
of a test particle with mass $m$ and charge $q$ in Minkowski space-time
whose spatial section has the nontrivial Chimney Euclidean topology $E_{11}$.
Curves for manifolds with a fixed $b=1$ and increasing compact lengths $a =1, 3, 6$,
are shown to roughly illustrate the topological Minkowskian limit for $E_{11}$
topology.
For these curves the greater is the value of the compact topological lengths
the longer is the time interval $\Delta t_{\mbox{ng}}$ for which the velocity
dispersion of is negligible.
In the topological Minkowskian limit $a \rightarrow \infty $ the whole space
becomes simply-connected, and one has $\Delta t_{\mbox{ng}} \rightarrow \infty $,
so during the whole time there is no velocity dispersion 
of the charged particle in Minkowski space-time with simply-connected spatial
topology.
\label{E_11z-a124}  }
\end{minipage}
\end{center}
\end{figure}

Two out of the nine orientable Euclidean three-dimensional topologies have two
compact directions.
In this section we study in more details the stochastic motions of charged particles
in Minkowski space-time whose spatial section has the Chimney space topology $E_{11}$.
This $3-$space is globally homogeneous. Thus the elements $\gamma$ of the group $\Gamma$
are translations. The fundamental domain is an open parallelepiped (Chimney), and one
identifies the two pairs of faces through independent translations.
Taking  $x$ and $y$ as the two independent compact directions and $a$ and $b$ as
the associated compact lengths, the expression for the spatial separation is then
given in the second row of Table~\ref{Tb-Spatial-separation}.

Following a similar procedure to that employed  for the Slab topology in the previous section,
for the spatial section with Chimney $E_{11}$ topology the electromagnetic field correlation
function is given by equation~\eqref{electric2} with spatial separation given in
Table~\ref{Tb-Spatial-separation}. In the coincidence limit
$\mathbf{x} \rightarrow \mathbf{x}'$ one has $r^2 = n_x^2a^2 + n_y^2b^2$,
$\widetilde{r_x}^2 = -n_x^2a^2 + n_y^2b^2$, $\widetilde{r_y}^2 = n_x^2a^2 - n_y^2b^2\,$
and  $\widetilde{r_z}^2 = n_x^2a^2 + n_y^2b^2\,$.
The global homogeneity of $E_{11}$ topology implies that $A_i = B_i = 1$ for $i= x, y, z$.
Inserting these limiting relations 
into~\eqref{eqdispersion3}   
after some simplifying calculation we have that the 
components of the mean squared velocity dispersion are given by
\begin{align}   \label{x-Dispers-E11}
\langle \Delta v_x^2\rangle
 = \frac{q^2}{2\pi^2m^2}
&\sum\limits_{{\!n_y=-\infty}}^{{\infty\;\;'}}
\Biggl[ \,\,
\sum\limits_{{n_x=-\infty}}^{{\infty\;\;'}}
\left\{ \frac{t^2n_y^2b^2}{(n_x^2a^2 + n_y^2b^2)^2(t^2 - n_x^2a^2 - n_y^2b^2)}
\Biggr. \right.   \nonumber   \\
& \Biggl. \left. \qquad + \,\frac{t(-2n_x^2a^2 + n_y^2b^2)}{4(n_x^2a^2 + n_y^2b^2)^{5/2}} \,
\ln\left(
 \frac{(\sqrt{n_x^2a^2 + n_y^2b^2} - t)^2}{(\sqrt{n_x^2a^2 + n_y^2b^2} + t)^2} \right)
\right\}
\,\,\, \Biggr] \,,
\end{align}
$\langle \Delta v_y^2\rangle = \langle \Delta v_x^2\rangle$ with 
$n_y \rightleftarrows n_x$ and $b \rightleftarrows a$,
and
\begin{align}  \label{z-Dispers-E11}
\langle \Delta v_z^2\rangle
 = \frac{q^2}{2\pi^2m^2}
&\sum\limits_{{\!n_y=-\infty}}^{{\infty\;\;'}}
\Biggl[ \,\,
\sum\limits_{{n_x=-\infty}}^{{\infty\;\;'}}
\left\{    \frac{t^2}{(n_x^2a^2 + n_y^2b^2)(t^2 - n_x^2a^2 - n_y^2b^2)}
\Biggr. \right.   \nonumber   \\
& \Biggl. \left. \qquad \quad + \,   \frac{t}{4(n_x^2a^2 + n_y^2b^2)^{3/2}}
\ln\left( \frac{(\sqrt{n_x^2a^2 + n_y^2b^2} - t)^2}{(\sqrt{n_x^2a^2 + n_y^2b^2} + t)^2} \right)
\,      
\right\}
\,\,\,  \Biggr] \,.
\end{align}         

It should be noticed that only when the $3-$space manifold has equal compact lengths ($a=b$)
the associated $x$ and $y$ components of the dispersion have identical behavior.  In this
case the roles played by the compactness in these directions are equivalent. 
The velocity dispersion behavior in this specific Chimney manifold corresponds to the
solid line curves in Figs.~\ref{E_11xy-a124} and~\ref{E_11z-a124}.

From the equation for $\langle \Delta v_y^2\rangle$ along with
Eqs.~\eqref{x-Dispers-E11}, \eqref{z-Dispers-E11}  a first important outcome,
which is also made evident in Figs.~\ref{E_11xy-a124} and \ref{E_11z-a124} and
that ratifies previous results with the Slab topology, is that compactification of
two independent directions gives rise to stochastic motion of charged particles whose
resulting dispersion $\langle \Delta \mathbf{v}^2 \left(\mathbf{x}, t\right) \rangle$ has
components not only in the two compact directions but also in the noncompact
direction $z$.
The solid line curve Figs.~(\ref{E_11xy-a124}) and~(\ref{E_11z-a124}) for the spatial
manifold with compact lengths $a=b=1$ shows the divergences in the velocity dispersion,
which again is of topological origin. The other curves are for different manifolds
($a=3b=3$, $a=6b=6$) also contain topological divergence at $t \simeq3 $ and $t \simeq6$
not completely displayed in these figures that was produced to illustrate mainly other
features of the velocity dispersion.

To examine the topological Minkowskian limit as the compact length $a$ increases
we have fixed the value of the compact length $b=1$ and increased $a=1, 3, 6$, etc. 
The curves for three manifolds with increasing compact length $a$ are
also shown in Figs.~\ref{E_11xy-a124}  and~\ref{E_11z-a124}.
For these curves the greater is the value of the compact topological lengths
the longer is the time interval $\Delta t_{\mbox{ng}}$ for which the velocity
dispersion of is negligible.
Thus, in the  Minkowskian topological limit $a\,, b  \rightarrow \infty$ the whole
space is simply-connected, and one has $\Delta t_{\mbox{ng}} \rightarrow \infty $,
i.e. during the whole time there is no velocity dispersion 
of the charged particle in Minkowski space-time with simply-connected spatial
topology.
This outcome for Minkowski space-time with Chimney spatial manifolds extends the result
of the previous section with the Slab topology, and again it is consistent with an earlier result
in the literature that ensures that quantum vacuum fluctuations would not produce stochastic
motions of charged particles in Minkowski space-time with simply-connected spatial section.

\subsection{\bf  Chimney space with half turn -- $\mathbf{E_{12}}$ } \label{secChimney-hturn}

When the Euclidean distance $d(\mathbf{x}, \gamma \mathbf{x})$ between a
point $\mathbf{x} \in M_3 $ and its image $\gamma \mathbf{x}$ is a constant
for all points $\mathbf{x} \in M_3$,  then the holonomy $\gamma$ is a
translation. In globally homogeneous quotient manifolds all elements  $\gamma$
of the covering group $\Gamma$ are translations. In these manifolds all
points $\mathbf{x}$ are topologically indistinguishable. Thus, the effect of
the topology in the evolution of a physical system in globally homogeneous
manifold is expected to be the same regardless of spatial position in $M_3$.
This is precisely the case of Slab and Chimney spaces, which have been used in
sections~\ref{secSlab} and \ref{secChimney} as spatial
section of Minkowski space-time. 

In globally inhomogeneous manifolds the distance function
$d(\mathbf{x}, \gamma \mathbf{x})$ is not constant, and
the length of the closed geodesic associated with at least one
non-translational $\gamma$ depends on the point $\mathbf{x} \in M_3$.
In this way, in globally inhomogeneous manifolds the identification
of at least one pair of faces of the fundamental domain is made through
a combination of a rotation followed by a translation along an axis.
The Chimney with half turn  of Table~\ref{Tb-4-Orient-manifolds}
is a globally inhomogeneous manifold.

In this section we report the results of our study of the motions of charged
test particles in Minkowski space-time whose spatial sections are endowed with
the Chimney with half turn topology.  Our main goal here is
to concretely exhibit the dependency of the velocity dispersion with the
particle position in the spatial section endowed with $E_{12}$ topology

Following a similar procedure to that employed in the previous
sections~\ref{secSlab} and \ref{secChimney} the electromagnetic field
correlation function is given by equation~\eqref{electric2} with spatial
separation given in Table~\ref{Tb-Spatial-separation}, where $x$ and $y$
were taken as the two independent compact directions and $a$ and $b$ are
the associated compact lengths.

Inserting the coincidence limit ($\mathbf{x} \rightarrow \mathbf{x}'$)
expressions for $r^2$, $\widetilde{r_x}^2$, $\widetilde{r_y}^2$ and
$\widetilde{r_z}^2$
along with $A_i$ and $B_i$  ($i= x, y, z$)
into~\eqref{eqdispersion3}   
after simplifying calculations we end up with
lengthy expressions for components of the mean squared velocity 
dispersion, which we do not present here for the sake of conciseness. These
expressions were used in the numerical calculations for the plots of the figures
in this section.

\begin{figure*}[tb]
\begin{center} 
\includegraphics[width=7.1cm,height=5.9cm]{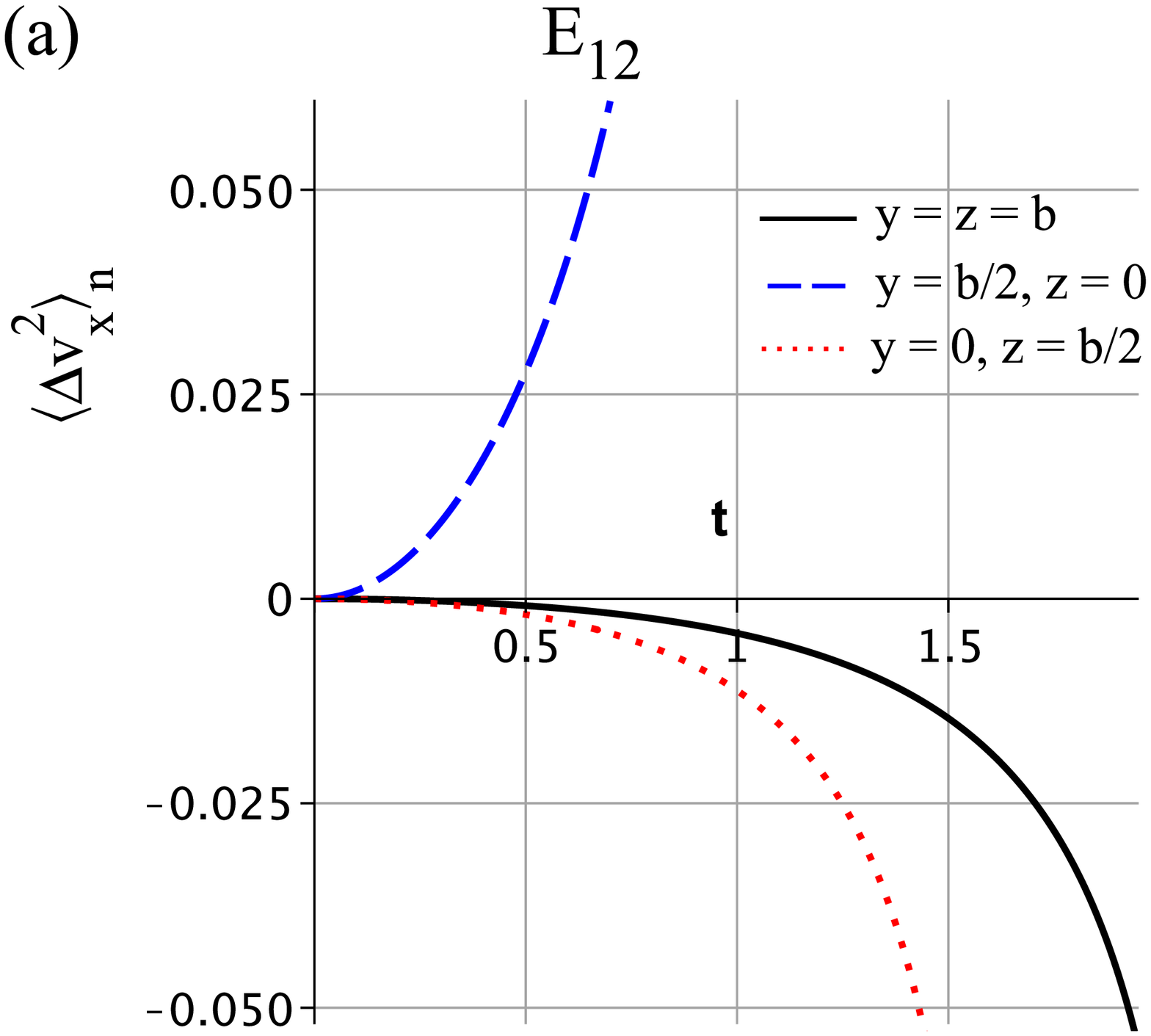}  
 \hspace{8mm}  %
\includegraphics[width=7.1cm,height=5.9cm]{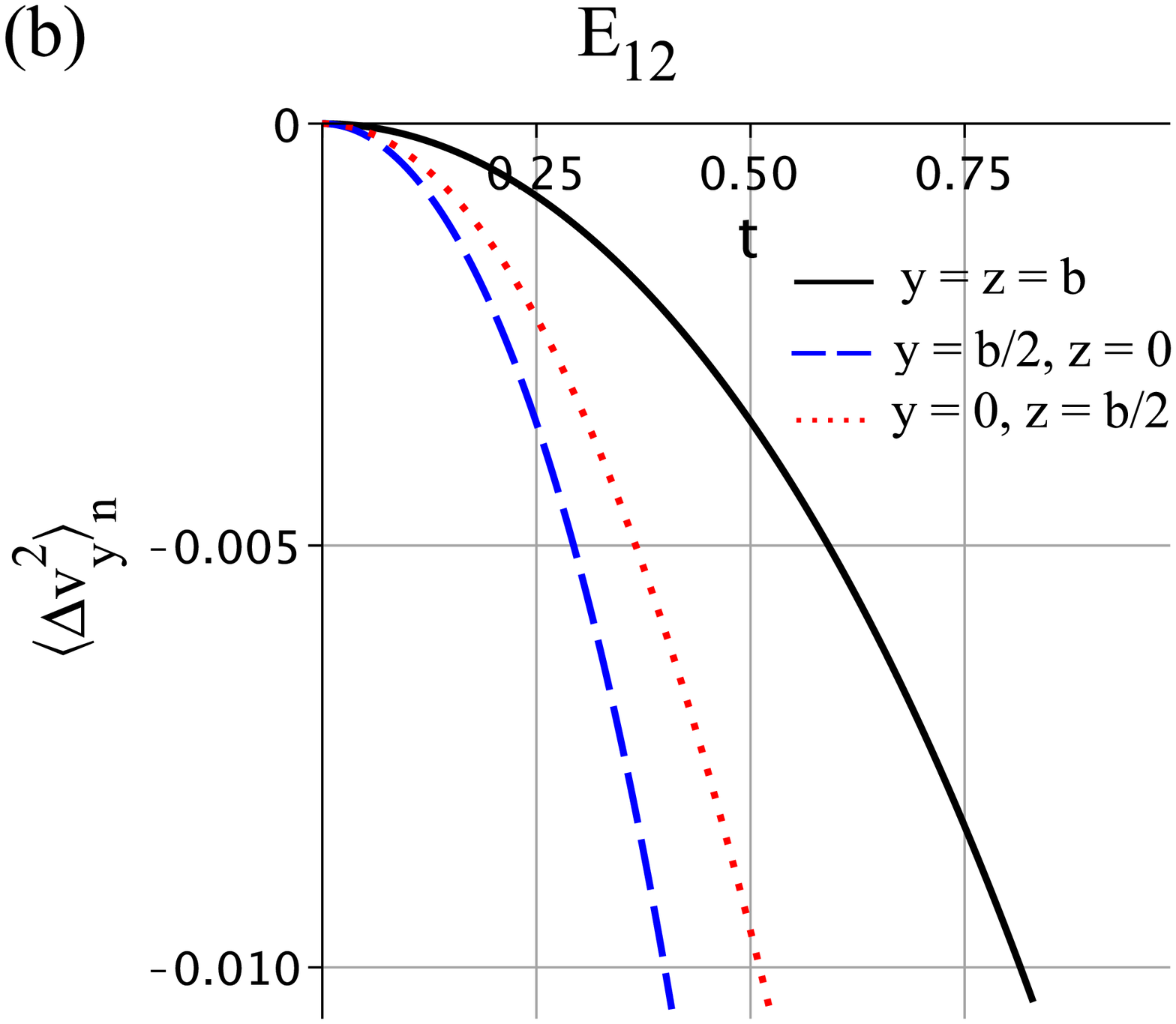}  
\begin{minipage}[t]{\textwidth}  \renewcommand{\baselinestretch}{0.96}  
\caption{The $x$  and  $y$  components [ panels ({\bf a})in the left hand side,
and ({\bf b}) in the right ] of the normalized velocity dispersion
$\langle{\Delta} \mathbf{v}^2
(\mathbf{x}, t) \rangle_{\,\mbox{n}}$ [Eq.~\eqref{normalized-disp}]
particle with mass $m$ and charge $q$ in Minkowski space-time whose spatial
section has the Euclidean nontrivial $E_{12}$ topology. Curves for a manifold
with equal compact length $a=b=1$ are depicted.
These curves illustrate that the time-evolution of the velocity dispersion of
the charged particle depends upon the particle's position in the space.
In each panel different curves correspond to different particle's positions
in the spatial section of Minkowski space-time. 
From the topological symmetry $z$ is a  noncompact direction with distinct
time evolution that is separately exhibited in Fig.~\ref{Fig_E_12z_abc1}. 
\label{Fig_E_12xy-abc1}  } .
\end{minipage}
\end{center}
\end{figure*}  

\begin{figure}[bt]
\begin{center} 
\includegraphics[width=7.1cm,height=5.9cm]{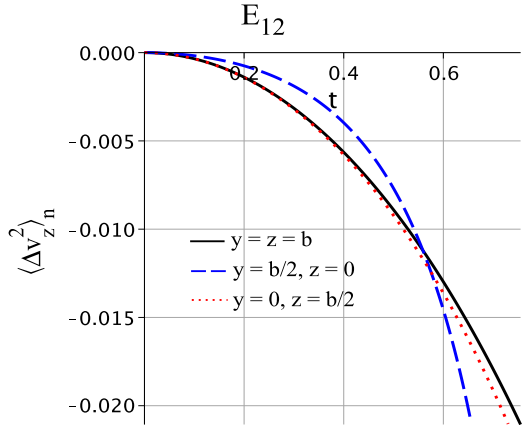} \vspace{-4mm}  
\begin{minipage}[t]{\textwidth}  \renewcommand{\baselinestretch}{0.96}
\caption{
The $z$  components of the normalized velocity dispersion
$\langle{\Delta} \mathbf{v}^2
(\mathbf{x}, t) \rangle_{\,\mbox{n}}$ [Eq.~\eqref{normalized-disp}]
of a test particle with mass $m$ and charge $q$ in Minkowski space-time
with spatial section endowed with the nontrivial Euclidean topology
$E_{12}$. Curves for a manifold with compact lengths $a =b=1$ are shown.
Different curves correspond to different particle's positions
in the spatial manifold. 
The curves illustrate that the time-evolution of the velocity dispersion
of the charged particle depends upon the particle's position in the
spatial section of Minkowski space-time. 
\label{Fig_E_12z_abc1}   }
\end{minipage}
\end{center}
\end{figure}

Panels ({\bf a}) and ({\bf b}) of Fig.~\ref{Fig_E_12xy-abc1} along with
Fig.~\ref{Fig_E_12z_abc1} show,
respectively, the time behavior of $i= x, y, z$  components of the
normalized mean squared velocity fluctuations 
$\langle{\Delta} \mathbf{v}^2 (\mathbf{x}, t) \rangle_{\,\mbox{n}} \equiv
(m^2/q^2)\,\langle\Delta \mathbf{v}^2\,(\mathbf{x}, t )\rangle$
of a test particle with mass $m$ and charge $q$ in Minkowski space-time
whose spatial section is a Chimney with half a turn manifold with
topological lengths $a=b=1$.
In the numerical calculation of each component in Fig.~\ref{Fig_E_12xy-abc1}
and Fig.~\ref{Fig_E_12z_abc1} we have taken in the summations $100$ terms of
the topological contribution to the dispersion, namely $n_x \neq 0\,, n_y \neq  0$
in the interval $[-50,50]$.

Three different spatial positions of the test particle 
$y=z=b$, $y=b/2, z=0$,  and $y=0, z=b/2$ have been taken in order to make
evident the distinct time evolution of the normalized dispersions for each
particle position. Figures~\ref{Fig_E_12xy-abc1} and~\ref{Fig_E_12z_abc1}
show the different time behavior of the  components of the
dispersion, including the  divergences for the velocity dispersions
that now also depend on the particle position.  
Preliminary numerical calculations for $E_{12}$ indicate that
for the above particle's positions the curves for the components of the
velocity dispersion  present, in the Minkowskian limit, similar pattern
of the corresponding curves for the other homogeneous manifolds.

An important outcome from these figures is that the time evolution of
the dispersion component depends on the test particle position
in the spatial section $M_3$. In other words, the effect of the topology
depends on the spatial position of the test particle.
This suggest that the time evolution of a physical system can
in principle be used to unveil the global homogeneity or inhomogeneity
which are important topological properties of the space-time.

\subsection{\bf  $3\,-$Torus \ -- \  $E_{1}$}  \label{3-torus}

Possibly the best known example of three-dimensional Euclidean space with
nontrivial topology is the $3-$Torus family of globally homogeneous
manifolds with three independent compact directions.
The manifolds in the $3-$Torus class have therefore higher degree
of connectedness than those in the homogeneous Slab
($E_{16}$) and Chimney ($E_{11}$) families which we have considered
in the previous sections.

To include an entirely compact Euclidean manifolds in our analysis  we
have also examined the motions of charged particles in Minkowski space-time
in which the spatial section is endowed with the $3-$Torus topology.
However, given the previous results concerning the time evolution
in Minkowski space-time with the Slab ($E_{16}$) and Chimney ($E_{11}$)
spatial topologies, it is expected that the $3-$Torus spatial topology
will clearly give rise to  stochastic motion of charged test particles.
In this way, our main point here is not to verify that these expected motions
can indeed occur, but rather to show that the previous outcomes obtained 
for Chimney, Slab and the simply-connected Minkowskian limit can be
recast as appropriate limits in the $3-$Torus spatial topology.

In general, the fundamental cells for $3-$Torus topology can be taken to be
a parallelepiped with different compact lengths $a, b, c$, whose faces are
identified through independent translations to form the manifolds.

We again follow a procedure similar to that used in sections~\ref{secSlab},
\ref{secChimney} and  \ref{secChimney-hturn} but now the electromagnetic field
correlation function is given by equation~\eqref{electric3} with spatial separation
$r$ given by the fourth entry in Table~\ref{Tb-Spatial-separation}, where $x$, $y$
and $z$ are the three independent compact directions, and $a$, $b$ and $c$ are
the associated compact lengths.
In the coincidence limit $\mathbf{x} \rightarrow \mathbf{x}'$ one has the spatial
separation reduces to $r^2 = n_x^2a^2 + n_y^2b^2 + n_z^2c^2$. The other terms $\widetilde{r_i}$
that are needed to have the components of the velocity dispersion~\eqref{eqdispersion4}
in the coincidence limit are given by
$\widetilde{r_x}^2 = -n_x^2a^2 + n_y^2b^2 + n_z^2c^2\,$, \
$ \widetilde{r_y}^2 = n_x^2a^2 - n_y^2b^2 + n_z^2c^2\,$,
and  $\widetilde{r_z}^2 = n_x^2a^2 + n_y^2b^2- n_z^2c^2$. \
As for the constants $A_i$ and $B_i$ we have $A_i = B_i = 1$ for $i = x, y, z$ from
the global homogeneity of the $3-$Torus.

Inserting these coincidence limiting relations
into~\eqref{eqdispersion4} after simplifying calculations we end up with
lengthy expressions for the components of the  velocity
dispersion, which we do not present here for conciseness, but
that were used in the numerical calculations in the plots of the
figures we discuss in what follows.

\begin{figure*}[tb]
\begin{center}
\includegraphics[width=7.1cm,height=5.9cm]{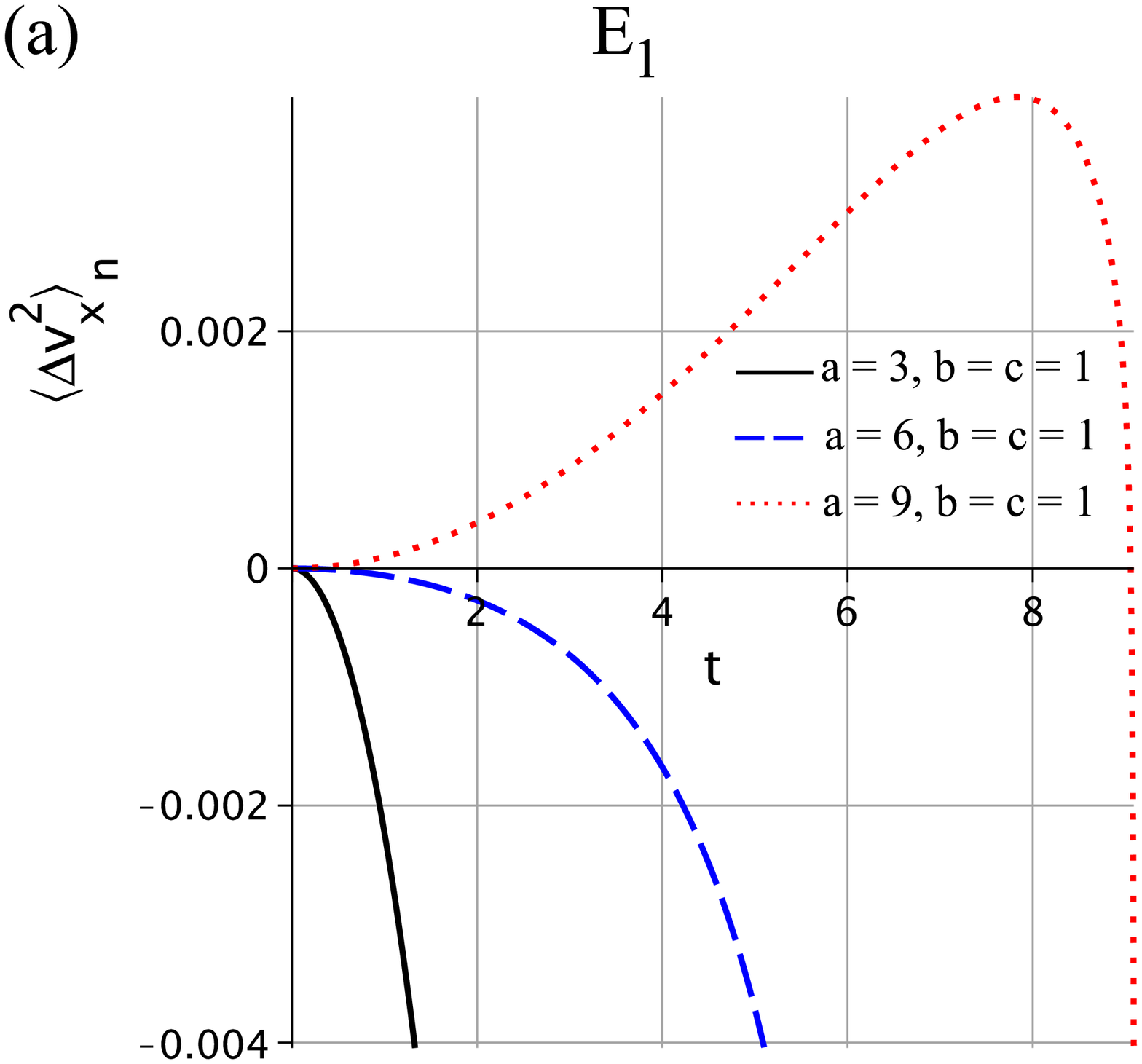}    
 \hspace{8mm}  
\includegraphics[width=7.1cm,height=5.9cm]{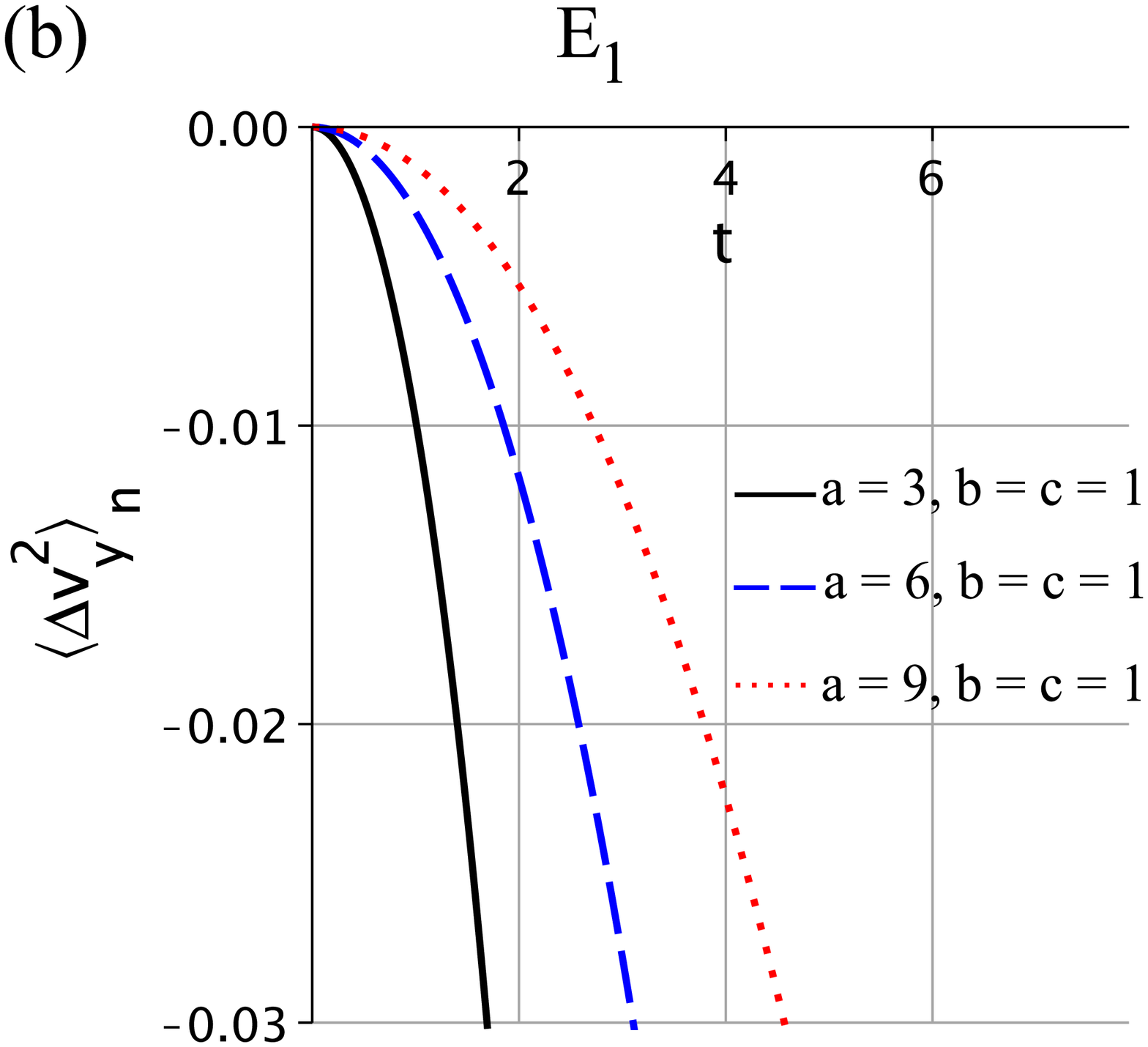}    
\begin{minipage}[t]{\textwidth}  \renewcommand{\baselinestretch}{0.96}  
\caption{The $x$  and  $y$  components [ panels ({\bf a}) on the left, and ({\bf b}) on
the right hand side ]
of the normalized velocity dispersion
$\langle{\Delta} \mathbf{v}^2 (\mathbf{x}, t) \rangle_{\,\mbox{n}} \equiv
(m^2/q^2)\,\langle\Delta \mathbf{v}^2\,(\mathbf{x}, t )\rangle$ of a test
particle with mass $m$ and charge $q$ in Minkowski space-time whose spatial
section has the $3-$Torus topology, $E_{1}$. Curves for manifolds
with equal compact length $b=c=1$ and increasing $a=3,6,9$  are shown to illustrate
the topological limit for small time $t$.
\label{Fig-E1xy-a369-bc1}   } .
\end{minipage}
\end{center}
\end{figure*}  

\begin{figure}[bt]
\begin{center} 
\includegraphics[width=8cm,height=5.9cm]{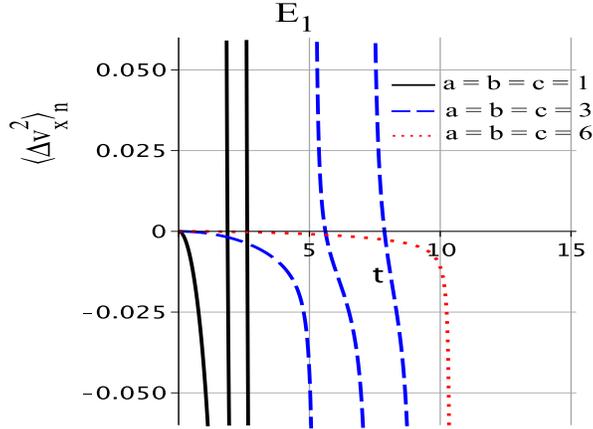}   
\begin{minipage}[t]{\textwidth}  \renewcommand{\baselinestretch}{0.96}
\caption{
The $x$ component of the normalized velocity dispersion
$\langle{\Delta} \mathbf{v}^2 (\mathbf{x}, t) \rangle_{\,\mbox{n}} \equiv
(m^2/q^2)\,\langle\Delta \mathbf{v}^2\,(\mathbf{x}, t )\rangle$
of a test particle with mass $m$ and charge $q$ in Minkowski space-time
with spatial has the $3$-Torus topology $E_{1}$. The $y$ and $z$ components
time evolution of the normalized dispersion are equal to the $x$ component
since $a=b=c$. Curves for manifolds with compact lengths $a=b=c=1,3,6$
are shown.
The divergent behavior of the normalized velocity dispersion for some values
of the time $t$ arises from periodic conditions imposed by $E_{1}$ topology
on the covering space $\mathbb{E}^3$.
The divergent behavior of the normalized velocity dispersion
arises from periodic conditions imposed by the topology.
\label{Fig_E1x-abc136}   }
\end{minipage}
\end{center}
\end{figure}

Panels ({\bf a}) and ({\bf b}) of Fig.~\ref{Fig-E1xy-a369-bc1}
show, respectively, the time evolution of the $x$ and $y$ components
of the normalized dispersion
$\langle{\Delta} \mathbf{v}^2 (\mathbf{x}, t) \rangle_{\,\mbox{n}}$
of a charged test particle in Minkowski space-time whose spatial sections
are manifolds endowed with a $3-$Torus topology.

To make use of our physical system in the approach to the Chimney space
$E_{11}$ as a limiting of $3-$Torus when two topological lengths are
fixed and a remaining length increases,
we have taken $b=c=1$ and increasing values $a=3, 6, 9$ in panels
{\bf a} and {\bf b} of  Fig.~\ref{Fig-E1xy-a369-bc1}. Clearly for
this choice of lengths the $z$ component of the velocity dispersion
has the same time evolution of $y$ component, which is exhibited
in the panel {\bf b}.
In the numerical calculation of each component in of Figs.~\ref{Fig-E1xy-a369-bc1}
and Fig.~\ref{Fig_E1x-abc136} as in the previous plots we have taken
$100$ terms in the summations, i.e.  $n_x \neq 0\,, n_y \neq  0\,, n_z \neq 0 $
in the interval $[-50,50]$.
The divergences in the component of the velocity dispersion in this figures come
from the $3$ compactification of the Torus. This was not the main features
we illustrate in these figures, though.

The comparison of  Fig.~\ref{Fig-E1xy-a369-bc1}
with Figs.~\ref{E_11xy-a124} 
for the Chimney space make apparent that the evolution of our physical
system in Minkowski space-time with Chimney spatial topology $E_{11}$
can be recovered from the evolution of the system in Minkowski with
a $3-$Torus topology.

Finally, to the extent that the greater is the value of the compact
lengths ($a,b,c$) the longer is the time interval in which  the components
of the dispersion are negligible, the Fig.~\ref{Fig_E1x-abc136}
indicates that the topological Minkowskian limit can  also be obtained
from the $3-$Torus when all lengths grow indefinitely, i.e.
$(a\,, b\,, c)  \rightarrow \infty$. Also compare, for increasing
the compact lengths, Fig.~\ref{Fig_E1x-abc136} 
with  Fig.~\ref{E_11xy-a124} and then examine Fig.~\ref{E16xy_Minkows}
for small time $t$.
\section{Final Remarks} \label{Con-Rem} 

In the standard  approach to model 
the physical world,  besides the points on the space-time manifold representing the
physical events, we have fields that satisfy appropriated differential
equations,  which  locally express the physical laws.
In the framework of field theories in curved space-times it is often 
assumed that a geometry, solution of the gravitational field equations,
couples to other fields and constrains their dynamics.
In this geometric approach the role played by the topology is either
implicitly (or explicitly) neglected or left open. 

Although they might not be recognized at first sight, topological
considerations are sometimes essential in physics (see
related Refs.~\cite{BGRT-1998,GRTB-2000,MMS-2015,RTT-1998},
and references therein quoted).
In quantum field theory, which handles, e.g.,  with field dynamics and its
fluctuations on differentiable manifolds, questions as to what
extent and how physical outcomes depend upon, are induced, or even
driven by a nontrivial topology, are very important.

Quantum vacuum fluctuations of the electromagnetic field in the
standard Minkowski space-time with simply-connected spatial section
seem not to produce observable effects on the motion of a charged test
particle.   
However,  when \textit{changes} in the background space
for the fluctuations are carried out, as for example by
the insertion of plane-boundaries into the three-dimensional space, %
the resulting mean squared velocity of a test charged particle is 
not null~\cite{yf04,yc04}. 

In this article we have tackled the question as to whether a nontrivial
spatial topology of Minkowski space-time provides conditions  
for a charged particle to undergo stochastic motion when subjected to
fluctuations of the electromagnetic field in empty space.
To answer this question, we have derived the mean squared velocity 
dispersion 
of a  charged test particle 
in Minkowski flat space-time whose spatial section has
one, two and three independent compact directions.
Equations~\eqref{eqdispersion2}~--~\eqref{eqdispersion4} explicitly
give the velocity dispersion in these cases with compact spatial
orientable topologies. In these equations
the spatial separation $r$ takes different forms
(Table~\ref{Tb-Spatial-separation}) so as to  capture the
periodic conditions imposed by the spatial topology.
In brief, equations~\eqref{eqdispersion2}~--~\eqref{eqdispersion4}
show that quantum vacuum fluctuations in an empty Minkowski
space-time with nontrivial spatial topology allow for stochastic
motions of charged particles.

To further explore the role played by the spatial topology in the
evolution of the velocity dispersion, we have concretely examined
the details of the  motion of a charged particle
for the $3-$spaces endowed with four flat
topologies of Table~\ref{Tb-4-Orient-manifolds},
namely three globally homogeneous Slab ($E_{16}$), Chimney ($E_{11}$),
$3-$Torus ($E_{1})$, and the globally inhomogeneous Chimney with
half a turn ($E_{12}$) topology.
A general outcome from the study with these topologies is that
compactification in just one direction is sufficient to produce
motion of charged particles with velocity dispersion components
in the compact and noncompact directions.
Here, differently from the motion obtained via the insertion of 
planes, the  motion of a charged particle under vacuum fluctuations
in Minkowski space-times occurs with no change of the smooth manifold
attributes of the space.
Also, for the globally homogeneous spatial topologies $E_{1}$,
$E_{11}$ and $E_{16}$ the global spatial homogeneity is unaltered
from the outset. In this way, the topological effects on the
whole evolution of the velocity dispersion is the same regardless
of the particle's position in the spatial section $M_3$.

It is well-known that Minkowski space-time is spatially flat and
locally homogeneous. Local homogeneity is a geometrical characteristic
of metric manifolds.
However, in dealing with topological spaces we have global homogeneity
and inhomogeneity of topological nature (Section~\ref{TopSet}).
An interesting question that arises in this context is whether a local experiment
can be prepared to possibly unveil these global (topological) properties of
the space.  
To illustrate that the motion of a charged particle under vacuum fluctuations can
potentially be used to capture or detect global inhomogeneity, we have also examined
the motion of charged particle in Minkowski space-time with $E_{12}$ spatial
globally inhomogeneous topology.
Figures~\ref{Fig_E_12xy-abc1} and~\ref{Fig_E_12z_abc1} show that different spatial
position $\mathbf{x}$ of the particle 
leads to different curves for the velocity dispersion. Hence, the time evolution
of the velocity dispersion for a charged particle under electromagnetic
fluctuations can be locally used to unveil the global inhomogeneity of the space.

It seems to be unsettled whether electromagnetic quantum vacuum fluctuations
in standard Minkowski space-time with simply-connected spatial section would 
allow for motion of a charged test particle. This would be an observable effect of
vacuum fluctuations.
In this paper we have also tackled this issue by considering the topological Mikowskian limit,
in which the infinite spatial manifold with simply-connected topology can be obtained
from globally homogeneous manifolds through limit when the compact topological lengths
grow indefinitely. We have found that, regardless the globally homogeneous topology
we start from (seed manifold),  
the greater is the value of the topological length the longer is the time interval
$\Delta t_{\mbox{ng}}$ for which the velocity dispersion is negligible.
In the limit when the topological lengths
$(a\,, b\,, c)$ tend to infinite, the time interval $\Delta t_{\mbox{ng}}$
in which the dispersion is negligible also tends to infinite,
$\Delta t_{\mbox{ng}} \rightarrow \infty $.  Thus during the
whole time there is no motion of a charged particle in
Minkowski space-time with simply-connected spatial topology.
In Fig.~\ref{E16xy_Minkows}, for example,  we illustrate this topological
limit for small time $t$, increasing compact length $a$ and for manifolds
with the Slab spatial topology $E_{16}$.
This result holds for Mikowskian limits of other manifolds with homogeneous
topologies $E_{11}$ and $E_1$ (Figs.~\ref{E_11xy-a124}, \ref{E_11z-a124}
and~\ref{Fig-E1xy-a369-bc1} and ~\ref{Fig_E1x-abc136}).
These topological Minkowskian limit results make apparent that no motion of a
charged particle arises from quantum electromagnetic fluctuations in the
standard Minkowski space-time with simply-connected spatial section.
The ultimate motive behind the stochastic motion of a charged particle under
electromagnetic quantum vacuum fluctuations is the nontrivial space topology.

\begin{acknowledgments}

M.J. Rebou\c{c}as acknowledges the support of FAPERJ under a CNE E-26/202.864/2017 grant,
and thanks CNPq for the grant under which this work was carried out.
C.H.G. Bessa is funded by the Brazilian research agency CAPES. 
We are also grateful to V.B. Bezerra for motivating discussions and
A.F.F. Teixeira for reading the manuscript and indicating
omissions and misprints.
\end{acknowledgments}


\end{document}